\documentclass[12pt]{iopart}
\usepackage[scr=boondoxo,scrscaled=1.05]{mathalfa}
\usepackage{tikz}
\usetikzlibrary{calc,decorations.markings}
\usepackage{epsfig}
\usepackage{graphicx}
\usepackage{rotating}
\usepackage{amssymb}
\usepackage{placeins}
\expandafter\let\csname equation*\endcsname\relax
\expandafter\let\csname endequation*\endcsname\relax
\usepackage{amsmath}
\usepackage[a4paper]{geometry}
\usepackage{subcaption}
\usepackage{float}   
\usepackage[font=small,labelfont=bf,justification=justified,format=plain]{caption}
\usepackage[active]{srcltx}
\usepackage{url}
\usepackage{hyperref}
\usepackage{centernot}
\usepackage{dsfont,bm}
\usepackage{xcolor}
\usepackage{pgfplots}
\usepackage{stackengine}
\usepackage{enumitem}
\usepackage{siunitx} 
\usepackage[scr=boondoxo,scrscaled=1.05]{mathalfa}
\hypersetup{
	colorlinks   = true, 
	urlcolor     = blue, 
	linkcolor    = blue, 
	citecolor   = red 
}
\colorlet{darkred}{red!55!black}
\colorlet{darkgreen}{green!25!black}
\usepackage{placeins}
\usepackage{tikz}
\usepackage{tkz-euclide}
\usetikzlibrary{calc,decorations.markings}
\usepackage{times,graphicx,xcolor}
\usepackage{delarray,fancybox}
\usepackage{mathdots}
\usepackage{eurosym}
\usepackage{relsize}
\usepackage{soul}
\usepackage{cancel}

\colorlet{darkred}{red!55!black}
\colorlet{darkgreen}{green!25!black}

\newcommand{\tf}{t_{\mathfrak{f}}}
\newcommand{\smi}[1]{\scriptscriptstyle{(#1)}}
\newcommand{\smp}[2]{\scriptscriptstyle{(#1:#2)}}
\newcommand{\tim}[1]{\mathscr{t}_{\mathfrak{#1}}}
\newcommand{\timt}{t}
\providecommand{\keywords}[1]{\textbf{\textit{Keywords:}} #1}

\begin{document}

	\title{Optimal control of engineered swift equilibration of nanomechanical oscillators}

	\author{Julia Sanders and Paolo Muratore-Ginanneschi
	}
	
	\address{University of Helsinki, Department of Mathematics and Statistics
		P.O. Box 68 FIN-00014, Helsinki, Finland}
    \ead{julia.sanders@helsinki.fi}	
	\ead{paolo.muratore-ginanneschi@helsinki.fi}

    \begin{abstract}
        We propose a reformulation of the problem of optimally controlled transitions in stochastic thermodynamics. We impose that any terminal cost specified by a thermodynamic functional should depend only on state variables and not on control protocols, according to the canonical Bolza form. In this way, we can unambiguously discriminate between transitions at minimum dissipation between genuine equilibrium states, and transitions at minimum work driving a system from a genuine equilibrium to a non-equilibrium state. For underdamped dynamics subject to a mechanical force, genuine equilibrium means a Maxwell-Boltzmann probability distribution defining a vanishing current velocity. Transitions at minimum dissipation between equilibria are a model of optimal swift engineered equilibration. Remarkably, we show that transitions at minimum work do not directly imply explicit boundary conditions on terminal values of parameters of the mechanical force and on control protocols. Thus, the problem often discussed in the literature, that optimal protocols need terminal jumps to satisfy boundary conditions, completely disappears. The quantitative properties of optimal controls are entirely determined by the form of the penalty modelling an experimental setup. 
More generally, we use centre manifold theory to analytically account for the tendency of optimal controls to exhibit a \emph{turnpike property}: optimal protocols in the bulk of the control horizon tend to converge to a universal centre manifold determined only by the running cost. Exponential deviations from the centre manifold occur at the ends of the control horizon in order to satisfy the boundary conditions. Our findings are supported numerically.

For illustration, our analysis is performed on the experimentally relevant case of nano-oscillators, but extension of our approach to feedback control in the presence of anharmonic forces is possible, albeit at the price of a considerably increased computational cost.    \end{abstract}

    
    {
		\let\newpage\relax    
		\maketitle
	}

	\section{Introduction}

	Existing nano-manipulation techniques mean that machines can be made that operate accurately at length scales of the order of \SIrange{10}{100}{\nm}, e.g. \cite{MaRoDiPePaRi2016,MaPeGuTrCi2016,AlLaJu2020,GoBaAsNoQuRo2021,DaCiBe2023}. 
    Real-time measurement and control of thermodynamic transitions between target states in time horizons of \SIrange{10}{100}{\micro\second} has been demonstrated in multiple experimental setups e.g. \cite{CuMaPeGuTrCi2016,BaCiBe2024,RaGuGuOdTr2023}. 
	These developments motivate interest in finding protocols to robustly steer a nano-machine to accomplish precise tasks while satisfying criteria of thermodynamic efficiency.
	Optimal control theory \cite{LibD2012, BecJ2021} provides a natural mathematical framework to find efficient protocols. In particular, if efficiency is defined by minimisation of the entropy production, optimal feedback control relates the second law of thermodynamics 
	\cite{AuGaMeMoMG2012} for open classical systems obeying a Langevin-Smoluchowski (overdamped) dynamics and optimal mass transport \cite{VilC2009}.
	
    In many applications, protocols are needed to implement fast transitions between equilibrium states \cite{ChCiGuOdTr2018,LiQuTu2017,FrZhChMaDeWe2021,GuJaPlPrTr2023,BaWhCiBe2024}. One example is in the design of a heat nano-engine that can reach the thermodynamic limit of efficiency at finite cycle time, by means of protocols that reproduce the same output as in a quasistatic process \cite{AlLaJu2020}. More generally, transitions between equilibrium end-states avoid the onset of a relaxation dynamics outside of the control horizon. Stability of the end-states is particularly desirable to improve power requirements for monitoring mesoscopic chemical or biological processes such as those reported in \cite{CoRiJaSmTiBu2005}.
	
	\subsection{Optimal vs Engineered Swift Equilibration Protocols}
	In much of the existing literature, optimal and engineered swift equilibration protocols rely on the highly stylised hypothesis that instantaneous manipulations of the mechanical potential are possible. Although this hypothesis leads to welcome mathematical simplifications, it also introduces some inherent shortcomings that prevent the model from consistently describing some relevant aspects of actual physical processes. We devote a few words here to motivate our claim.
    
	  To start with, it is important to emphasize the conceptual difference between the optimal control and the engineered swift equilibration control models. Both optimal and engineered swift equilibration aim at constructing a thermodynamic transition defined by the assignment of system state indicators at each end of the control horizon. For feedback optimal control (closed-loop; where the control is dependent on the current system state), this transition can be formulated as a generalised Schr\"odinger bridge problem \cite{ChMGSc2021}, in which the end values of the probability density of the open system are the boundary data.
	 	 Assuming all necessary smoothness, the probability density of the open system obeys a Fokker-Planck equation.

    Optimal feedback control of generalised Schr\"odinger bridges minimises an assigned average thermodynamic cost. As the dynamics is modelled by a Markov process, optimisation is attained by minimising the instantaneous average cost rate \cite{LibD2012,BecJ2021}. The cost is a function of the system phase space coordinates and satisfies a dynamic programming equation. The Fokker-Planck and dynamic programming equations are first order in time and are coupled by a stationary condition on the drift that does not involve time derivatives.
    The solution is thus fully specified by the assignment of \emph{two} boundary conditions in time: the end values of the system probability density. 
    These boundary condition cannot generically describe equilibria, even if they do take the form of Maxwell-Boltzmann distributions. 
    Requiring that the probability distributions at the end of the control horizon be equilibria (stationary solutions of the Fokker-Planck equation) entails assigning additional boundary conditions in time, the vanishing of the current velocity, that would \emph{overdetermine} the problem. In the literature, we often encounter the suggestion that one should associate discontinuities at the end of the control horizon to optimal protocols solving otherwise already well-posed problems to satisfy the equilibrium requirement. 
    Accordingly, the mechanical potential {\textquotedblleft}jumps{\textquotedblright} to the value needed for the probability distributions at each end of the control horizon to be in equilibrium, \emph{without affecting} the cost of the transition or the probability distribution of the system. 
    
    Mathematically, the end-of-horizon discontinuity idea may be upheld by invoking 
    generalised criteria for existence of solutions to differential equations (see e.g. \cite[\S~3.1.3]{LibD2012}).     
    Physically, the jump idea is an attempt to model the dynamics on time-scales that are not resolved in the formulation of the generalised Schr\"odinger bridge problem. It is hardly, however, a tenable solution for several reasons. As pointed out in \cite{PlPrTrCuOd2021}, assuming a jump is of little help from the experimental point of view, as it does not translate into any well-defined operational protocol. In addition, we argue that the self-consistence of this idea is questionable. Underdamped, and with greater reason overdamped, dynamics model a classical open system under the assumption of a strong separation of time scales with those of the environment, subsumed in a white noise force \cite{PavG2014}. 
      Within this framework, it is questionable to justify temporal variations of the system's drift on scales faster than white noise as elements of admissible control protocols. Similar concerns regarding the consistence of control protocols with the derivation of the dynamics from microscopic models are often raised in quantum control theory \cite{ScCaPeSuAn2015}. 
	Further doubts are also fuelled by the reformulation of the problem in terms of jump processes. The corresponding optimal protocols admit as scaling limits those of the overdamped dynamics only under additional fine-tuning assumptions \cite{MGMePe2012}.

	 The upshot of this discussion is that optimal control protocols are adapted either to establish universal lower bounds as for Landauer's principle \cite{LeOrPoSn2019}, or to describe a physical process in which equilibration of the end states is not a requirement. This is the case, for example, when transitions constitute branches of a Stirling's engine cycle as in \cite{ScSe2008,MGSc2015}. 
     
     Within the control horizon, engineered swift equilibration protocols only need to satisfy the Fokker-Planck equation. Hence the requirement that assigned boundary conditions at each end of the control horizon are equilibria does not produce an overdetermined problem. Engineered swift equilibration protocols are then constructed based on 
    the simplicity of their implementation \cite{GuRuKiToMa2019}. Whilst this is of practical advantage, it opens the question of how to conceptualise optimal transitions between equilibria. More precisely, the question is how to resolve the neglected time scales in generalised Sch\"odinger bridges. 

    \subsection{Motivation of the present work}
    \label{sec:motivation}
   
	The motivation of this work is to address the mathematical and physical difficulties discussed above. From the mathematical point of view, our perspective is that boundary conditions should be imposed on system state variables rather than controls \cite{LibD2012} to avoid overdetermining an optimal control problem already at the formulation level. In other words, problems are well posed when they are amenable to the canonical Bolza form of optimal control theory \cite[\S~3.3.2]{LibD2012}.  
    From the physical point of view, implementing protocols without bandwidth limitations may cause 
    signal distortions in high-accuracy experiments of quantum control \cite{HiCa2018}.  As a step towards a more realistic model, 
    we instead consider the microscopic mechanical potential as an element of the system state  onto which boundary conditions can be imposed. Correspondingly, we  identify controls with (the parameters of) a force governing the variation in time of the mechanical potential acting on the open system. Here, we mainly conceptualise this force as the result of macroscopic manipulations whose cost may depend on the setup. 
	
	In order to simplify the discussion, we focus on the dynamics of a nano-oscillator in one dimensional configuration space. In other words, we restrict the attention to open systems whose probability distribution under the action of a quadratic mechanical potential is Gaussian at any time. Although restrictive, this assumption corresponds to many  experimental setups \cite{MaRoDiPePaRi2016,MaPeGuTrCi2016,ChBeGuOdTrPeCi2018,AlLaJu2020,RaGuGuOdTr2023,BaCiBe2024}.
	We note that when dealing with Gaussian states, the distinction between feedback and open-loop control (where the control is exerted by means of parameters independent of the current system state) is somewhat blurred, as the set of controls is finite. 
	
	\subsection{Structure of the work and main findings}
	
	The structure of this work is the following. In Section~\ref{sec:dynamics}, we define the nanooscillator control model. Drawing from \cite{MaPeGuTrCi2016}, we suppose that it is possible to experimentally determine whether at the beginning and at the end of the control horizon the system is in equilibrium. Correspondingly, we specify the state of the system by assigning first and second order cumulants \emph{and} the mechanical force on the system at both ends of the control horizon. We identify controls with the time derivative of the 
    parameters of the mechanical force. 
    
    In Section~\ref{sec:td}, we show that 
    we can unambiguously discriminate engineered swift equilibration transitions at minimal dissipation from transitions connecting target states by performing a minimum amount of thermodynamic work on the system. In this latter case, target states are described by probability distributions that are not necessarily in equilibrium with the microscopic mechanical force exerted on an open system. This distinction also allows us to characterise the relaxation to equilibrium process that may eventually take place after the end of the control, which is done in Section~\ref{sec:relax}. 
    
    In Section~\ref{sec:formulation}, we describe the set of admissible controls and use Pontryagin’s maximum principle to formulate first order optimality conditions for engineered swift equilibration at minimum dissipation and minimum work transitions in Sections~\ref{sec:MP} and~\ref{sec:ext}. From a physics point of view, the most natural formulation is to identify the cost of a transition with the work done on the system. Controls are then restricted to a compact set. However, due to difficulties we discuss in implementing this numerically, we describe how self-concordant barriers can be used in their place. Self-concordant barriers are used as a tool to apply methods of singular perturbation theory \cite{FenN1979,GuKu2009,CaTe2017} for reduction of first order conditions for optimality to a universal normal form in Section~\ref{sec:centre}. In Section~\ref{sec:od} we describe their overdamped limit in the bulk of the control horizon.  
   
   In Section~\ref{sec:num}, we contrast numerical solutions of the equations expressing first order conditions for optimality  
    (indirect optimisation) and their normal form obtained in Section~\ref{sec:centre} with numerical methods of direct optimisation \cite{WacA2009,TaCoWaLa2019}. 
    
    The last section is devoted to conclusions and outlook.

\section{Open system dynamics}
\label{sec:dynamics}

We consider a one-dimensional underdamped linear dynamics in non-dimensional variables
\begin{equation}
	\begin{split}
		&	\mathrm{d}\mathscr{q}_{t}=\varepsilon \,\mathscr{p}_{t} \mathrm{d}t  
		\\
		&	\mathrm{d}\mathscr{p}_{t}=-\left( \mathscr{p}_{t}-\varepsilon\,\mathscr{u}_{t}+\varepsilon\,\mathscr{k}_{t}\,\mathscr{q}_{t}\right)\mathrm{d}t+\sqrt{2}\,\mathrm{d}\mathscr{w}_{t}
	\end{split}
	\label{dyn:ud}
\end{equation}
where  $\varepsilon$ denotes the order parameter of the overdamped expansion. We relate non-dimensional variables to dimensional ones by identifying
\begin{align}
	\varepsilon=\sqrt{\frac{\tau^{2}}{m\,\beta\,\ell^2}}
	\nonumber
\end{align}
where $\tau$ is the Stokes time, $m$ the mass of the system, $\beta$ the inverse temperature of the environment, and $\ell$ the typical length scale of the problem. 

In (\ref{dyn:ud}), the drift on momentum increments is the sum of a linear friction term and the gradient of the quadratic potential 
\begin{align}
	U_{t}(q)=\frac{\mathscr{k}_{t}\,(q -\mathscr{c}_{t})^{2}}{2}
	\nonumber
\end{align}
characterised by a time-dependent stiffness $\mathscr{k}_{t}$ and stationary point $ \mathscr{c}_{t}$. The product of these quantities 
\begin{align}
	\mathscr{u}_{t}=\mathscr{k}_{t}\,\mathscr{c}_{t}
	\nonumber
\end{align}
determines the position independent term in (\ref{dyn:ud}). In what follows, we find expedient to regard $\mathscr{u}_{t}$ as independent dynamical varible.

\subsection{Control equations}

We assume that stiffness and centre obey the first order differential equations
\begin{equation}
	\begin{split}
		&	\dot{\mathscr{k}}_{t}=\lambda_{t}
		\\
		& \dot{\mathscr{u}}_{t}=\gamma_{t}
	\end{split}
	\label{dyn:ceqs}
\end{equation}
We identify the pair of deterministic time dependent functions $(\lambda_{t},\gamma_{t})$ with the control quantities of the dynamics. We require the controls to be essentially bounded \cite{BoSiSu2021}. Physically, (\ref{dyn:ceqs}) states that the microscopic mechanical potential only varies in response of external actions with delay, modelled by the solution of differential equations.

\subsection{Cumulant dynamics}
\label{sec:cd}

As well known, the transition probability of a system of linear stochastic differential equations with additive noise is Gaussian at any time. Correspondingly, the Fokker-Planck equation reduces to a solvable finite hierarchy of differential equations for first and second order cumulants. Hence, if we define 
\begin{align}
	&
	\begin{split}
	&	\mathscr{x}^{\smi{1}}_{t}=\operatorname{E}\left(\mathscr{q}_{t}-\mathscr{x}^{\smi{5}}_{t}\right)^{2}
	\\
	&	\mathscr{x}^{\smi{2}}_{t}=\operatorname{E}\left(\mathscr{p}_{t}-\mathscr{x}^{\smi{6}}_{t}\right)\left(\mathscr{q}_{t}-\mathscr{x}^{\smi{5}}_{t}\right)
	\\
	&	\mathscr{x}^{\smi{3}}_{t}=\operatorname{E}\left(\mathscr{p}_{t}-\mathscr{x}^{\smi{6}}_{t}\right)^{2}
	\\
		&	\mathscr{x}^{\smi{4}}_{t}=\mathscr{k}_{t}
	\end{split}
	&&\mbox{and} &
	\begin{split}
			&\mathscr{x}^{\smi{5}}_{t}=	\operatorname{E}\mathscr{q}_{t}
		\\
		&\mathscr{x}^{\smi{6}}_{t}=	\operatorname{E}\mathscr{p}_{t}
		\\
	&	\mathscr{x}^{\smi{7}}_{t}=	\mathscr{u}_{t}
	\end{split}
	\nonumber
\end{align}
then the set of cumulant equations complemented by the control equations (\ref{dyn:ceqs})
\begin{align}
	\begin{split}
		&		\dot{\mathscr{x}}_{t}^{\smi{1}}=2 \,\varepsilon\,  \mathscr{x}_{t}^{\smi{2}}	
	\\
	&		\dot{\mathscr{x}}_{t}^{\smi{2}}=- \mathscr{x}_{t}^{\smi{2}}-\varepsilon \left(\mathscr{x}_{t}^{\smi{4}}\mathscr{x}_{t}^{\smi{1}} -\mathscr{x}_{t}^{\smi{3}}\right)  
	\\
	&		\dot{\mathscr{x}}_{t}^{\smi{3}}=2 \left(1-\mathscr{x}_{t}^{\smi{3}}-\varepsilon \,\mathscr{x}_{t}^{\smi{4}}\, \mathscr{x}_{t}^{\smi{2}} \right)		
	\\
	&		\dot{\mathscr{x}}_{t}^{\smi{4}}= \lambda_{t} 			
	\end{split}
	\label{dyn:2cum}
\end{align}
and
\begin{align}
	\begin{split}		
		&		\dot{\mathscr{x}}_{t}^{\smi{5}}=\varepsilon\,\mathscr{x}_{t}^{\smi{6}}
		\\
		&		\dot{\mathscr{x}}_{t}^{\smi{6}}= -\mathscr{x}_{t}^{\smi{6}}+\varepsilon\left(\mathscr{x}_{t}^{\smi{7}}-\mathscr{x}_{t}^{\smi{4}} \,\mathscr{x}_{t}^{\smi{5}} \right)
		\\
		&		\dot{\mathscr{x}}_{t}^{\smi{7}}=\gamma_{t} 
	\end{split}
	\label{dyn:1cum}
\end{align}
fully characterise the statistics of (\ref{dyn:ud}). In particular, the system of equations (\ref{dyn:2cum}) governing the evolution of second order cumulants is independent of the system (\ref{dyn:1cum}). In an ideal experiment, it is thus possible to control an open system with Gaussian statistics by squeezing or spreading the distribution while holding the mean value fixed or, conversely, by shifting the mean value without affecting the average size of the fluctuations around it. 

\subsection{Gaussian boundary conditions for the control problem}

We are interested in transitions between Gaussian distributions of Maxwell-Boltzmann form at both ends of a control horizon $[0\,,\tf]$. 
We thus always require
\begin{align}
	\mathscr{x}_{0}^{\smi{2}}=\mathscr{x}_{\tf}^{\smi{2}}=\mathscr{x}_{0}^{\smi{6}}=\mathscr{x}_{\tf}^{\smi{6}}=0
		\label{bc:MB}
\end{align}
meaning that the position-momentum cross correlation and the mean value of the momentum process vanish at both ends of the control horizon. We identify the open system temperature with the variance of the momentum process. Hence, a unit value at both ends of the control horizon
\begin{align}
	\mathscr{x}_{0}^{\smi{3}}=\mathscr{x}_{\tf}^{\smi{3}}=1
	\label{bc:isoth}
\end{align}
entails that the system is at the same temperature as the surrounding bath at the beginning and end of the transition.

 In order to fully specify a transition governed by (\ref{dyn:ud}) and (\ref{dyn:ceqs}), we need to assign the value of the mechanical force acting on the system at both ends of the control horizon. Concretely, we need to assign $\mathscr{x}_{t}^{\smi{4}}$, and $\mathscr{x}_{t}^{\smi{7}}$ at
$t=0,\tf$. This is mathematically equivalent to the specification of the probability current \cite{NelE2001}. 
When the Gaussian Maxwell-Boltzmann conditions (\ref{bc:MB}), (\ref{bc:isoth}) hold, engineered swift equilibration transitions correspond to the requirements
\begin{align}
	&	\mathscr{x}_{0}^{\smi{1}}=\frac{1}{\mathscr{x}_{0}^{\smi{4}}} && 	\mathscr{x}_{\tf}^{\smi{1}}=\frac{1}{\mathscr{x}_{\tf}^{\smi{4}}} 
	\label{bc:sc2}
\end{align}
and
\begin{align}
	&	\mathscr{x}_{0}^{\smi{5}}=\frac{\mathscr{x}_{0}^{\smi{7}}}{\mathscr{x}_{0}^{\smi{4}}} && \mathscr{x}_{\tf}^{\smi{5}}=\frac{\mathscr{x}_{\tf}^{\smi{7}}}{\mathscr{x}_{\tf}^{\smi{4}}} 
	\label{bc:sc1}
\end{align}
In this framework, transitions between or towards non-equilibrium states are those where the boundary conditions \eqref{bc:sc2} and \eqref{bc:sc1} are not satisfied.

\section{Work done on the system as a thermodynamic cost functional}
\label{sec:td}

The average work done on the system during a stochastic thermodynamic transition in the horizon $[0\,,\tf]$ is \cite{PePi2020}
\begin{align}
	\mathcal{W}_{\tf}=\operatorname{E}\int_{0}^{\tf}\mathrm{d}s\,(\partial_{s}U_{s})(\mathscr{q}_{s})
	\nonumber
\end{align}
where the partial derivative only acts on the explicit time dependence of the mechanical potential $ U$.
A straightforward application of It\^o's lemma shows that, for any sufficiently regular but not necessarily harmonic mechanical potential, the mean work is amenable to the form
\begin{align}
	\mathcal{W}_{\tf}=\operatorname{E} \left(U_{\tf}(\mathscr{q}_{\tf})-U_{0}(\mathscr{q}_{0})+\frac{\mathscr{p}_{\tf}^{2}
		-\mathscr{p}_{0}^{2} }{2}\right)+\operatorname{E}\int_{0}^{\tf}\mathrm{d}t\,\mathscr{p}_{t}^{2}-\tf
	\nonumber
\end{align}
According to the first law of thermodynamics, the mean heat released by the system is 
\begin{align}
	\mathcal{Q}_{\tf}=	\mathcal{W}_{\tf}-\operatorname{E} \left(U_{\tf}(\mathscr{q}_{\tf})-U_{0}(\mathscr{q}_{0})+\frac{\mathscr{p}_{\tf}^{2}
		-\mathscr{p}_{0}^{2} }{2}\right)
	\nonumber
\end{align}
By Clausius' law, the mean heat released 
is the entropy variation of the environment in non-dimensional units. The total entropy variation during a transition is the sum of the entropy variation of the environment and the system. The system entropy variation is specified by the Gibbs-Shannon entropy of the
probability distribution $\mathrm{f}_{t}$ of the system
\begin{align}
	\mathcal{S}_{t}= -\int_{\mathbb{R}^2}\mathrm{d}p\mathrm{d}q\,\mathrm{f}_{t}(p,q)\ln \mathrm{f}_{t}(p,q)=-\operatorname{E}\ln \mathrm{f}_{t}(\mathscr{p}_{t},\mathscr{q}_{t})
\nonumber
\end{align}
We thus arrive at the expression of the second law of thermodynamics
\begin{align}
	\mathcal{E}_{\tf}=\mathcal{S}_{\tf}-\mathcal{S}_{0}+\mathcal{Q}_{\tf}=\operatorname{E}\int_{0}^{\tf}\mathrm{d}t\,\Big{(}\mathscr{p}_{t} +\partial_{\mathscr{p}_{t}}\ln \mathrm{f}_{t}(\mathscr{p}_{t},\mathscr{q}_{t})\Big{)}^2\geq\,0
\label{td:ep}
\end{align}
which holds in this form for any mechanical potential \cite{PePi2020}. We refer to \cite{ChGa2008} for the derivation of (\ref{td:ep}) from fluctuation relations.

When the system dynamics is linear and the probability distributions at both ends of the control horizon are Gaussian and of Maxwell-Boltzmann type , i.e.(\ref{bc:MB}) and (\ref{bc:isoth}) hold true, we can further exhibit the dependence of the work on the cumulants
\begin{align}
	\mathcal{W}_{\tf}&=
\frac{\mathscr{x}_{\tf}^{\smi{4}}}{2}\left (\mathscr{x}^{\smi{1}}_{\tf}+\left (\mathscr{x}^{\smi{5}}_{\tf} - \frac{\mathscr{x}_{\tf}^{\smi{7}}}{\mathscr{x}_{\tf}^{\smi{4}}}\right )^{2}\right )-\frac{\mathscr{x}_{0}^{\smi{4}}}{2}
\left (\mathscr{x}^{\smi{1}}_{0}+\left (\mathscr{x}^{\smi{5}}_{0} - \frac{\mathscr{x}_{0}^{\smi{7}}}{\mathscr{x}_{0}^{\smi{4}}}\right )^{2}\right )
\nonumber\\
&+\int_{0}^{\tf}\mathrm{d}t\,\left(\mathscr{x}^{\smi{3}}_{t}+\mathscr{x}^{\smi{6}}_{t}{}^{2}\right)-\tf
	\label{td:work}
\end{align}
When we control the transition via (\ref{dyn:ceqs}), all quantities appearing in the boundary terms of work are state variables of the system. In other words, (\ref{td:work}) is the total transition cost sum of a \emph{running cost} produced by the integral over the average entropy production rate and a \emph{terminal cost} 
in standard \emph{Bolza form} (see e.g. \cite[\S~3.3.2]{LibD2012}).
We can therefore unambiguously discriminate between at least two classes of transitions. 
\begin{enumerate}[label=\textbf{C.\Roman*}]
	\item \label{C1}Transitions between equilibrium states: in such a case the average work coincides with the mean heat release $\mathcal{Q}_{\tf}$ :
	\begin{align}
	\mathcal{W}_{\tf}=	\mathcal{Q}_{\tf}=\int_{0}^{\tf}\mathrm{d}t\,\left(\mathscr{x}^{\smi{3}}_{t}+\mathscr{x}^{\smi{6}}_{t}{}^{2}\right)-\tf
		\label{work:ep}
	\end{align}
    The corresponding value of the mean entropy production is
    \begin{align}
	\mathcal{E}_{\tf}=	\mathcal{Q}_{\tf}+\frac{1}{2}\ln \frac{\mathscr{x}^{\smi{1}}_{\tf}}{\mathscr{x}^{\smi{1}}_{0}}\ge 0
    \label{work:ep2}
	\end{align}
	We readily see that (\ref{work:ep}), (\ref{work:ep2}) vanish if the system is maintained at Maxwell-Boltzmann equilibrium
	\begin{align}
 &   \begin{cases}
 \mathscr{x}^{\smi{1}}_{t}=\mathrm{const.} 
\\
	\mathscr{x}^{\smi{3}}_{t}=1 & \&  \hspace{0.8cm}\mathscr{x}^{\smi{6}}_{t}=0 
		\end{cases}
        & \forall\,t\,\in\,[0,\tf]
        \nonumber
	\end{align}
	Transitions in this class are models of optimal engineered swift equilibration. 
	\item \label{C2}Minimum work transitions to a target state
	\begin{align}
		\mathcal{W}_{\tf}=
		\frac{\mathscr{x}_{\tf}^{\smi{4}}}{2}\left (\mathscr{x}_{\tf}^{\smi{1}}+\left (\mathscr{x}_{\tf}^{\smi{5}} - \frac{\mathscr{x}_{\tf}^{\smi{7}}}{\mathscr{x}_{\tf}^{\smi{4}}}\right )^{2}\right )-\frac{1}{2}
		+\int_{0}^{\tf}\mathrm{d}t\,\left(\mathscr{x}^{\smi{3}}_{t}+\mathscr{x}^{\smi{6}}_{t}{}^{2}\right)-\tf
		\label{work:terminal}
	\end{align}
\end{enumerate}
The upshot is that, when thermodynamic transitions connect genuine equilibrium states, work optimisation is equivalent to the minimisation of mean heat release. In addition, when the control is exerted via (\ref{dyn:ceqs}), transitions at \emph{minimum work} (\ref{work:terminal}) become clearly distinct from those minimising the entropy production. 

\section{Formulation of the optimal control problems}
\label{sec:formulation}

The work functional (\ref{td:work}) is not convex in the controls $\lambda_{t}$ and $\gamma_{t}$. In order to ensure that the optimisation problem is well posed, we need either to refine the space of admissible controls or to introduce a convex penalty on protocols that can be enacted to steer the dynamics of the system.

\subsection{Penalty on admissible controls}
\label{sec:penalty}

We model the total cost of controlling the transition as
\begin{align}
	\mathcal{C}_{\tf}=\mathcal{W}_{\tf}+\int_{0}^{\tf}\mathrm{d}t\, \left(g\,V(\lambda_{t})+\tilde{g}\,\tilde{V}(\gamma_{t})\right)
	\label{penalty:general}
\end{align}
The {\textquotedblleft}penalty parameters{\textquotedblright} $g$ and $\tilde g$ couple a cost term
weighing the controls exerted on the mechanical potential felt by the nano-particle to the thermodynamic functional.
We consider three cases qualitatively describing analogous physical setups, but with mathematical properties that may be more adapted either to numerical or analytical inquiry.

Because the dynamics of the second order cumulants fully decouples from that of the first order cumulants, 
 we restrict attention to transitions that only change the position variance of the system. 

\subsubsection{Hard penalty}

We suppose that admissible controls are time dependent functions that are confined to a compact set, i.e.  satisfying for any $t\in [0\,,\tf]$
\begin{align}
	&	|\lambda_{t}|\,\leq\,\Lambda\,<\infty  \label{penalty:hard}
    \end{align} 
	
Finite dimensional approximations reduce the minimisation of a cost functional to finding the minima of a multidimensional function on a compact set. Weierstrass extreme value theorem \cite[\S~1.2.1]{LibD2012} then guarantees the existence of such minima. 
Physically, (\ref{penalty:hard}) directly models constraints on the intensity of controls that admit straightforward motivation in actual experimental setups: for instance constraints on the power range of an optical device used to confine a colloidal particle \cite{BaBePlRaTr2024}.
Furthermore, the assumption (\ref{penalty:hard}) is appealing, because as long as the controls satisfy (\ref{penalty:hard}), 
the optimisation problem is well posed by setting 
\begin{align}
	V(\lambda_{t})=0 
	\nonumber
\end{align}
in (\ref{penalty:general}). Correspondingly, the cost function preserves the universal form prescribed by stochastic thermodynamics.

\subsubsection{Logarithmic penalty}

Direct methods of optimal control \cite{RaoA2009,TeLiYuRe2021} approximate the total cost by means of a multidimensional objective function. Optimisation is thus recast into a minimisation problem to be tackled by non-linear programming. In such a setup, constraints such as (\ref{penalty:hard}) are {\textquotedblleft}softened{\textquotedblright} by couching them into the form of potential barriers. Major developments in optimisation theory over the last thirty years led to the discovery \cite{KarN1984} and refinement \cite{NeNe1994,NeTo2008} of interior point methods, giving rise to algorithms capable of minimising a convex programming problem in polynomial time. The key ingredient is the adoption of {\textquotedblleft}self-concordant{\textquotedblright} barriers to model constraints. In one dimension, the defining properties of these barriers are:
\begin{enumerate}
	\item the second derivative of the barrier is a Lipschitz function with universal constant equal to $2$ 
	\begin{align}
		\left |\frac{\mathrm{d}^{3}f}{\mathrm{d} x^{3}}(x)\right |\,\leq\,2 \left(\frac{\mathrm{d}^{2}f}{\mathrm{d} x^{2}}(x)\right)^{3/2}
		\label{penalty:scf}
	\end{align} 
	 \item the first derivative be a Lipschitz function with tunable constant $\vartheta$:
	 \begin{align}
	 	\left |\frac{\mathrm{d}f}{\mathrm{d} x}(x)\right |\,\leq\,\vartheta^{1/2} \left(\frac{\mathrm{d}^{2}f}{\mathrm{d} x^{2}}(x)\right)^{1/2}
	 	\label{penalty:scb}
	 \end{align} 
\end{enumerate}
Any univariate function satisfying (\ref{penalty:scf}) is called self-concordant. An objective constrained by self-concordant barriers attains a unique absolute minimum depending on the penalty parameter. As this latter varies, the minimum treads a {\textquotedblleft}central path{\textquotedblright} asymptotically tending to the solution of  the hard barrier problem. Self-concordance ensures that numerical computation of the limit takes polynomial time. 

Having in mind the above considerations, a suitable choice for the numerical analysis of (\ref{penalty:general}) is the self-concordant potential 
\begin{align}
	V(z)=
	\begin{cases}
		-\ln \left(1-\dfrac{z^{2}}{\Lambda^{2}}\right)	&  |z|\,\leq\,\Lambda
		\\[0.3cm]
		+\infty& |z|\,>\,\Lambda
	\end{cases}
	\label{penalty:log}
\end{align}
Complementary to direct methods are indirect methods based on the solution of extremal equations that optimal control necessarily satisfy. The Legendre transform of the barrier then enters the formulation of Pontryagin's maximum principle determining the extrema equations. For (\ref{penalty:log}) we find
\begin{align}
	\arg\max_{z  \in \mathbb{R}}\left(y\,z -g\,V(z)\right)=\frac{g}{y}\left(\sqrt{1+\frac{\Lambda^{2}\,y^{2}}{g^{2}}}-1\right)
	\nonumber
\end{align}

\subsubsection{Harmonic penalty}

Many qualitative features of optimal control problems subject to a {\textquotedblleft}hard penalty{\textquotedblright} can be captured by considering a harmonic model of penalty such as
\begin{align}
    V(z)=\Bigl(\frac{z}{\Lambda}\Bigr)^{2}
	\label{penalty:harmonic}
\end{align}
which also enjoys the self-concordant property. A harmonic potential approximates (\ref{penalty:log}) near the minimum i.e. for control actions that have a low impact on the total cost (\ref{penalty:general}).  In Section~\ref{sec:centre}, we use \eqref{penalty:harmonic} to apply centre manifold analysis to derive the universal form of the first order conditions for optimal control problem associated to the cumulant dynamics equations (\ref{dyn:2cum}), (\ref{dyn:1cum}).

\subsection{Effective cost}

Based on the discussion in Section~\ref{sec:td} above, we set out to analyse the following mathematical problems.

\subsubsection{\ref{C1}: Transitions between equilibrium states.}
\label{sec:ep}

The quantity to minimise is 
\begin{align}
	\mathcal{C}_{\tf}=\int_{0}^{\tf}\mathrm{d}t \,\left(\mathscr{x}_{t}^{\smi{3}}+g\,V(\lambda_{t})\right)
	\label{ep:cost}
\end{align}
where $g=0$ if the admissible control is subject to the constraint (\ref{penalty:hard}) and $g>0$ if instead we penalise large values of the controls with a confining potential such as (\ref{penalty:log}) or (\ref{penalty:harmonic}) 

The cost functional only consists of a time integral and is thus in Lagrange form \cite[\S~3.3.2]{LibD2012}. 
To apply Pontryagin's maximum principle in its most general form, we write the total cost in the equivalent Mayer form. 
\begin{align}
	\mathscr{x}_{t}^{\smi{0}}=\int_{0}^{t}\mathrm{d}s \,\left(\mathscr{x}_{s}^{\smi{3}}+g\,V(\lambda_{s})\right)
	\nonumber
\end{align}
and define
\begin{align}
	\tilde{\mathcal{C}}_{\tf}=\mathscr{x}_{\tf}^{\smi{0}}
	\nonumber
\end{align}
subject to the dynamical constraint 
\begin{align}
	\dot{\mathscr{x}}_{t}^{\smi{0}}=\mathscr{x}_{t}^{\smi{3}}+g\,V(\lambda_{t})
	\label{ep:Mayer}
\end{align}
in addition to (\ref{dyn:2cum}).

\subsubsection{\ref{C2}:  Transitions at minimum work.}
\label{sec:mw}

The cost functional is
\begin{align}
	\mathcal{C}_{\tf}=
	\frac{x_{\tf}^{\smi{4}}\,x_{\tf}^{\smi{1}}}{2}+\int_{0}^{\tf}\mathrm{d}s\,\left(x_{s}^{\smi{3}}+g\,V(\lambda_{s})\right)
	\label{mw:cost}
\end{align}
with an additional terminal cost now specified by the final value of the position correlation and of the stiffness of the microscopic mechanical potential felt by the system. 
Applying (\ref{ep:Mayer}), we get the equivalent Mayer form 
\begin{align}
	\tilde{\mathcal{C}}_{\tf}=
	\frac{x_{\tf}^{\smi{4}}\,x_{\tf}^{\smi{1}}}{2}+\mathscr{x}_{\tf}^{\smi{0}}
	\nonumber
\end{align}

\section{First Order Optimality Conditions}\label{sec:MP}

Pontryagin's maximum principle identifies first order, i.e. necessary, conditions for optimality versus {\textquotedblleft}needle{\textquotedblright} or McShane-Pontryagin variations of controls \cite[\S~4.2.3]{LibD2012}. It thus extends calculus of variations to trajectories that are also stationary with respect to discontinuous controls. To this end, the maximum principle applies to the Mayer form of a total cost functional. Once we impose the dynamics through Lagrange multipliers, we obtain an action functional of the form
\begin{align}
	\mathcal{A}_{\tf}=c\,\frac{x_{\tf}^{\smi{4}}\,x_{\tf}^{\smi{1}}}{2}+\mathscr{x}_{\tf}^{\smi{0}}+
	\int_{0}^{\tf}\mathrm{d}t\left(\sum_{\mathfrak{i}=\mathfrak{0}}^{\mathfrak{4}}\mathscr{y}_{t}^{\smi{i}}\dot{\mathscr{x}}_{t}^{\smi{i}}-\tilde{H}(\bm{\mathscr{x}}_{t},\bm{\mathscr{y}}_{t},\lambda_{t})\right)
	\label{MP:action}
\end{align}
where $\tilde{H}$ is the {\textquotedblleft}pre-Hamiltonian{\textquotedblright}
\begin{align}
	\tilde{H}(\bm{\mathscr{x}}_{t},\bm{\mathscr{y}}_{t},\lambda_{t})&= \mathscr{y}_{t}^{\smi{0}} \left(\mathscr{x}_{t}^{\smi{3}}+g \,V(\lambda_{t})\right)+2 \,\varepsilon\,\mathscr{y}_{t}^{\smi{1}}\,  \mathscr{x}_{t}^{\smi{2}} + 	\mathscr{y}_{t}^{\smi{2}}\left(- \mathscr{x}_{t}^{\smi{2}}-\varepsilon \left(\mathscr{x}_{t}^{\smi{4}}\mathscr{x}_{t}^{\smi{1}} -\mathscr{x}_{t}^{\smi{3}}\right)  \right) 
	\nonumber\\
	& 	+2\,\mathscr{y}_{t}^{\smi{3}}\, \left(1-\mathscr{x}_{t}^{\smi{3}}-\varepsilon \,\mathscr{x}_{t}^{\smi{4}}\, \mathscr{x}_{t}^{\smi{2}} \right)	+\mathscr{y}_{t}^{\smi{4}}\lambda_{t}
	\label{MP:pH}
\end{align}
and 
\begin{align}
c = \begin{cases}
    &0 \qquad \text{case \ref{C1}} \\
    &1 \qquad \text{case \ref{C2}}
\end{cases}
    \nonumber
\end{align}
An important consequence of  needle variations of the action functional in Mayer form (\ref{MP:action}) is that extremals can occur either when the Lagrange multiplier $\mathscr{y}_{t}^{\smi{0}}$ is a non-vanishing constant or equal to zero almost everywhere during the control horizon.  When $$\mathscr{y}_{t}^{\smi{0}}=0$$ almost everywhere, we encounter \emph{abnormal} extremals that are independent of the form of the cost functional. Hence, the first step to determine necessary conditions for optimality is usually to assess the existence of abnormal extremals \cite{BoSiSu2021}. From the physical point of view, the cost functional discriminates between distinct thermodynamic transitions. We thus expect optimal control to be determined by normal extremals corresponding to the condition
\begin{align}
	\mathscr{y}_{t}^{\smi{0}}=-1
	\label{MP:normal}
\end{align}
recovering the Lagrange (case \ref{C1}) or Bolza (case \ref{C2}) forms of the cost functional. We defer a proof of this statement to~\ref{sec:abnormal} and instead focus here on the analysis of normal extremals.

\subsection{Normal extremals}

Normal extremals obey the Hamilton equations 
\begin{align}
	&
	\begin{split}
&		\dot{\mathscr{x}}_{t}^{\smi{i}}=(\partial_{\mathscr{y}_{t}^{\smi{i}}}\tilde{H})(\bm{x}_{t},\bm{y}_{t},\lambda_{t})
		\\
&		 \dot{\mathscr{y}}_{t}^{\smi{i}}=-(\partial_{\mathscr{x}_{t}^{\smi{i}}}\tilde{H})(\bm{x}_{t},\bm{y}_{t},\lambda_{t}) 
	\end{split}
&&	 \mathfrak{i}=\mathfrak{1},\dots,\mathfrak{4}
	\label{MP:Heqs}
\end{align}
with the maximum condition
\begin{align}
	\lambda_{t}= \arg \max_{l \in \mathrm{A}} H(\bm{x}_{t},\bm{y}_{t}, l)
	\label{MP:argmax}
\end{align}
where $\mathrm{A}$ is the set of admissible controls. The terminal cost determines the boundary conditions of 
(\ref{MP:Heqs}), (\ref{MP:argmax}) that to fully specify extremals.

\subsubsection{\ref{C1}: Transitions between equilibrium states}
\label{MP:bc_ep}
The mean heat release (\ref{work:ep}) does not include any terminal cost.  The cumulants of the Gaussian Maxwell-Boltzmann 
distributions at the boundaries are assigned by (\ref{bc:MB}), (\ref{bc:isoth}). Transitions between equilibrium states also satisfy (\ref{bc:sc2}), which gives the boundary conditions for the stiffness of the mechanical drift. In such a case, the numerical value of the mean heat dissipation also coincides with that of the mean work. If boundary values of the stiffness violate (\ref{bc:sc2}),
the terminal states are not equilibria. This happens if we identify the control with the drift as done, for example, in \cite{AuGaMeMoMG2012,MGSc2014,SaBaMG2024}. 

\subsubsection{\ref{C2}: Transitions at minimum work.}
\label{MP:bc_mw}
We minimise the mean work (\ref{mw:cost}) 
while again assigning all cumulants of the Gaussian  Maxwell-Boltzmann end-of-horizon distributions in agreement (\ref{bc:MB}), (\ref{bc:isoth}). As in section~\ref{MP:bc_ep}, we assume that the system is in equilibrium at time $t=0$. In this case, the minimisation is also performed over the terminal value of the stiffness. Stationary variation then corresponds to the condition
\begin{align}
	\mathscr{y}_{\tf}^{\smi{4}}=-\dfrac{\mathscr{x}_{\tf}^{\smi{1}}}{2}
\label{noneq_bc}
\end{align}
We discuss the interpretation and consequences of this condition in section~\ref{sec:mwt} below.

\section{Analysis of regular extremals}
\label{sec:ext}
We now describe the explicit form of the stationary equations arising from the class of admissible controls and penalty potential.

\subsection{Normal extremals with controls subject to a convex penalty} If the penalty potential is a convex function of its argument like in (\ref{penalty:log}) or (\ref{penalty:harmonic}), enforcing the maximum condition (\ref{MP:argmax}) is equivalent to taking the Legendre transform of the pre-Hamiltonian with respect to the control \cite{BoPi2005}. The result is a Hamiltonian in the classical mechanics sense. Regardless of how we proceed, the optimal evolution necessarily satisfies the Hamiltonian system of differential equations
\begin{align}
	&		\begin{cases}
		\dot{\mathscr{x}}_{t}^{\smi{1}}=2 \,\varepsilon\,  \mathscr{x}_{t}^{\smi{2}}	
		\\
		\dot{\mathscr{x}}_{t}^{\smi{2}}=- \mathscr{x}_{t}^{\smi{2}}-\varepsilon \left(\mathscr{x}_{t}^{\smi{4}}\mathscr{x}_{t}^{\smi{1}} -\mathscr{x}_{t}^{\smi{3}}\right)  
		\\
		\dot{\mathscr{x}}_{t}^{\smi{3}}=2 \left(1-\mathscr{x}_{t}^{\smi{3}}-\varepsilon \,\mathscr{x}_{t}^{\smi{4}}\, \mathscr{x}_{t}^{\smi{2}} \right)		
		\\
		\dot{\mathscr{x}}_{t}^{\smi{4}}= b(\mathscr{y}_{t}^{\smi{4}})
	\end{cases}
	&& 
	\begin{cases}
		\dot{\mathscr{y}}_{t}^{\smi{1}}=\varepsilon\, \mathscr{y}_{t}^{\smi{2}} \,\mathscr{x}_{t}^{\smi{4}} 
		\\
		\dot{\mathscr{y}}_{t}^{\smi{2}}=-2\,\varepsilon\,\mathscr{y}_{t}^{\smi{1}}+\mathscr{y}_{t}^{\smi{2}}+2\,\varepsilon\, \mathscr{y}_{t}^{\smi{3}}  \,\mathscr{x}_{t}^{\smi{4}}
		\\
		\dot{\mathscr{y}}_{t}^{\smi{3}}=1-\varepsilon\, \mathscr{y}_{t}^{\smi{2}}+2\,\mathscr{y}_{t}^{\smi{3}}
		\\
		\dot{\mathscr{y}}_{t}^{\smi{4}}= 	\varepsilon\, \mathscr{y}_{t}^{\smi{2}}\,\mathscr{x}_{t}^{\smi{1}} +2\,\varepsilon \,\mathscr{y}_{t}^{\smi{3}}\, \mathscr{x}_{t}^{\smi{2}} 
	\end{cases}
	\label{ext:eqs}
\end{align}
The drift steering the stiffness is
\begin{align}
	b(y)=
	\begin{cases}
		\dfrac{g}{y}\left(\sqrt{1+\dfrac{\Lambda^{2}\,y^{2}}{g^{2}}}-1\right)
& \mbox{for the logarithmic penalty} \,(\ref{penalty:log})
		\\[0.4cm]
    \dfrac{\Lambda^2\, y}{2\,g}	&\mbox{for the harmonic penalty} \,(\ref{penalty:harmonic})
	\end{cases}
	\label{ext:Hvf}
\end{align}
We thus obtain a system with eight equations and eight boundary conditions determined either by section~\ref{MP:bc_ep} or  by \ref{MP:bc_mw}.

\subsection{Controls in a compact set with no penalty}
\label{sec:synthesis}

We now analyze (\ref{MP:action}) when $g$ in the pre-Hamiltonian is equal to zero and controls are confined in a compact set as by (\ref{penalty:hard}). In such a case optimal controls are generically brought about by contrasting  two classes of extremal conditions \cite{BoPi2005}.

\subsubsection{Controls on the boundary of the compact set}
\label{sec:push}

 The pre-Hamiltonian satisfies a stationary condition when evaluated  on controls belonging to the boundary of the compact set (\ref{penalty:hard}).
 Correspondingly
	\begin{align}
		\arg \max_{l \in \mathrm{A}} \tilde{H}(\bm{x}_{t},\bm{y}_{t}, l)=\operatorname{sgn}(\mathscr{y}_{t}^{\smi{4}}) \Lambda
		\label{push:bv}
	\end{align}
specifies the drift $b(\mathscr{y}_{t}^{\smi{4}})$ in (\ref{ext:eqs}).  An optimal control always satisfying (\ref{push:bv}) is commonly referred to as {\textquotedblleft}bang-bang{\textquotedblright}. 
	
\subsubsection{Controls stemming  from the differential stationary condition.}
\label{sec:noact}

It is still, however, possible that the pre-Hamiltonian is stationary on controls belonging to the interior of the admissible set. In such a case
	\begin{align}
		(\partial_{\lambda_{t}}H)(\bm{x}_{t},\bm{y}_{t},\lambda_{t})=0
		\label{noact:a1}
	\end{align} 
implies
	\begin{align}
		\mathscr{y}_{t}^{\smi{4}}=0
		\label{noact:c1}
	\end{align}
This is only an implicit condition on the control. The condition (\ref{noact:c1}) may only hold temporarily during the control horizon, as we discuss below. 

\subsubsection{Synthesis} In general, optimal paths are obtained by gluing trajectories obeying (\ref{push:bv}) and (\ref{noact:c1}) over sub-intervals of the control horizon. This operation is called \emph{synthesis} and is required when (\ref{noact:c1}) turns out to be more cost efficient than a pure bang-bang control. Yet, (\ref{noact:c1}) is not generically sufficient to ensure the fulfillment of the boundary conditions. To substantiate this last claim we proceed as in \cite{GoScSe2008,HegG2013,MGSc2017,BaBePlRaTr2024,LuMaPlSaBa2025}. In order for (\ref{noact:c1}) to hold during a finite time interval, it must be a conservation law of the Hamiltonian dynamics (\ref{MP:Heqs}) with control $\lambda_{t}$ so far unspecified. We obtain
\begin{align}
	0=	\dot{\mathscr{y}}_{t}^{\smi{4}}=	\varepsilon\, \mathscr{y}_{t}^{\smi{2}}\,\mathscr{x}_{t}^{\smi{1}} +2\,\varepsilon \,\mathscr{y}_{t}^{\smi{3}}\, \mathscr{x}_{t}^{\smi{2}} 
	\label{noact:a2}
\end{align}
As the position variance, $ \mathscr{x}_{t}^{\smi{1}} $, is non-vanishing (a condition always satisfied by smooth densities), we can therefore perform the following calculation. We find
\begin{align}
	\mathscr{y}_{t}^{\smi{2}}=-\frac{2\,\mathscr{y}_{t}^{\smi{3}}\, \mathscr{x}_{t}^{\smi{2}} }{\mathscr{x}_{t}^{\smi{1}}}
	\label{noact:c2}
\end{align}
The condition (\ref{noact:a2}) does not directly involve the control $\lambda_{t}$. Again, it can hold on a finite time interval if it is preserved by the dynamics. Upon differentiating it along trajectories satisfying (\ref{MP:Heqs}) and using (\ref{noact:c2}), we arrive at 
\begin{align}
	0=-2\,\varepsilon\,\mathscr{y}_{t}^{\smi{1}}\,\mathscr{x}_{t}^{\smi{1}} +2\,\mathscr{x}_{t}^{\smi{2}}+2\,\varepsilon\,\mathscr{y}_{t}^{\smi{3}}\,\mathscr{x}_{t}^{\smi{3}}
	\label{noact:a3}
\end{align}
 We solve for the Lagrange multiplier 
\begin{align}
	\mathscr{y}_{t}^{\smi{1}}=\frac{\mathscr{x}_{t}^{\smi{2}}+\varepsilon\,\mathscr{y}_{t}^{\smi{3}}\,\mathscr{x}_{t}^{\smi{3}}}{\varepsilon\,\mathscr{x}_{t}^{\smi{1}}}
	\label{noact:c3}
\end{align}
Again, (\ref{noact:a3}) does not yield any explicit condition on the control. We differentiate it again along trajectories satisfying (\ref{MP:Heqs}). Upon inserting (\ref{noact:c2}) and (\ref{noact:c3}) into the result, we get 
\begin{align}
	0	=\mathscr{x}_{t}^{\smi{1}} \left(\mathscr{x}_{t}^{\smi{2}}-2\,\varepsilon \, \left( \mathscr{x}_{t}^{\smi{3}}+  \,\mathscr{y}_{t}^{\smi{3}}\right)\right)+\varepsilon \, \mathscr{x}_{t}^{\smi{4}} \mathscr{x}_{t}^{\smi{1}}{}^2+2\, \varepsilon  \,
	\mathscr{x}_{t}^{\smi{2}}{}^2
	\label{noact:a4}
\end{align} 
Solving for the Lagrange multiplier, we get
\begin{align}
	\mathscr{y}_{t}^{\smi{3}}=\frac{1}{2}
	\left(\frac{\mathscr{x}_{t}^{\smi{2}}}{\varepsilon\, }+\frac{2 \mathscr{x}_{t}^{\smi{2}}{}^2}{\mathscr{x}_{t}^{\smi{1}}}-2 \mathscr{x}_{t}^{\smi{3}}+\mathscr{x}_{t}^{\smi{1}} \mathscr{x}_{t}^{\smi{4}}\right)
	\label{noact:c4}
\end{align}
Repeating this reasoning one last time: differentiating (\ref{noact:a4}), and using (\ref{noact:c2}), (\ref{noact:c3}), and (\ref{noact:c4}) yields
\begin{align}
	0=-\frac{8\, \varepsilon^2 \,\mathscr{x}_{t}^{\smi{2}}{}^3}{\mathscr{x}_{t}^{\smi{1}}}+\varepsilon\,  \mathscr{x}_{t}^{\smi{1}}{}^2 \left(\lambda_{t} -3 \,\mathscr{x}_{t}^{\smi{4}}\right)+2
	\varepsilon  \mathscr{x}_{t}^{\smi{2}} \left(4 \,\varepsilon  \,\mathscr{x}_{t}^{\smi{3}}-5 \mathscr{x}_{t}^{\smi{2}}\right)+\mathscr{x}_{t}^{\smi{1}} \left(9 \,\varepsilon  \mathscr{x}_{t}^{\smi{3}}-3 \mathscr{x}_{t}^{\smi{2}}-6 \varepsilon \right)
	\nonumber
\end{align}
 which can be finally solved for the control
 \begin{align}
 	\lambda_{t}=\frac{8 \,\varepsilon \, \mathscr{x}_{t}^{\smi{2}}{}^3}{\mathscr{x}_{t}^{\smi{1}}{}^3}+\frac{2 \,\mathscr{x}_{t}^{\smi{2}} \left(5 \,\mathscr{x}_{t}^{\smi{2}}-4 \,\varepsilon \, \mathscr{x}_{t}^{\smi{3}}\right)}{\mathscr{x}_{t}^{\smi{1}}{}^2}+\frac{3 \,\mathscr{x}_{t}^{\smi{2}}-9\,\varepsilon \, \mathscr{x}_{t}^{\smi{3}}+6\,\varepsilon }{\varepsilon \,\mathscr{x}_{t}^{\smi{1}}}+3 \mathscr{x}_{t}^{\smi{4}}
 	\nonumber
 \end{align}
  The upshot is that the stationary condition (\ref{noact:a1}) yields the differential system
 \begin{align}
 	\begin{split}
& 		\dot{\mathscr{x}}_{t}^{\smi{1}}=2 \,\varepsilon\,  \mathscr{x}_{t}^{\smi{2}}	
 		\\
& 		\dot{\mathscr{x}}_{t}^{\smi{2}}=- \mathscr{x}_{t}^{\smi{2}}-\varepsilon \left(\mathscr{x}_{t}^{\smi{4}}\mathscr{x}_{t}^{\smi{1}} -\mathscr{x}_{t}^{\smi{3}}\right)  
 		\\
& 		\dot{\mathscr{x}}_{t}^{\smi{3}}=2 \left(1-\mathscr{x}_{t}^{\smi{3}}-\varepsilon \,\mathscr{x}_{t}^{\smi{4}}\, \mathscr{x}_{t}^{\smi{2}} \right)		
 		\\
 &		\dot{\mathscr{x}}_{t}^{\smi{4}}= \dfrac{8 \,\varepsilon \, \mathscr{x}_{t}^{\smi{2}}{}^3}{\mathscr{x}_{t}^{\smi{1}}{}^3}+\dfrac{2 \,\mathscr{x}_{t}^{\smi{2}} \left(5 \,\mathscr{x}_{t}^{\smi{2}}-4 \,\varepsilon \, \mathscr{x}_{t}^{\smi{3}}\right)}{\mathscr{x}_{t}^{\smi{1}}{}^2}+\dfrac{3 \,\mathscr{x}_{t}^{\smi{2}}-9\,\varepsilon \, \mathscr{x}_{t}^{\smi{3}}+6\,\varepsilon }{\varepsilon \,\mathscr{x}_{t}^{\smi{1}}}+3 \mathscr{x}_{t}^{\smi{4}}
 	\end{split}
 	\label{noact:reqs}
 \end{align}
 complemented by the algebraic equations (\ref{noact:c1}), (\ref{noact:c2}), (\ref{noact:c3}), and (\ref{noact:c4}). The system (\ref{noact:reqs}) alone can only satisfy four out of the eight boundary conditions specifying an extremal, hence the need for synthesis.

\section{Centre manifold analysis}
\label{sec:centre}

Physical intuition suggests that the optimal control models in Section~\ref{sec:ext} describe qualitatively equivalent experimental setups. Our aim here is to quantitatively corroborate this intuition by studying the limit of vanishing $g$. 
To this end, we draw from the supplementary material of \cite{MGSc2014}. There, centre (also known as invariant) manifold analysis \cite{TiKoJo1994,HaKaKoOn1998} is used to identify universal properties of mean dissipation minimising protocols in underdamped transition between Gaussian states when the control is the mechanical potential.

The starting observation is that, as $g$ tends to zero, (\ref{ext:Hvf}) is not sensitive to the penalty mechanism as long as
\begin{align}
	\mathscr{y}_{t}^{\smi{4}} \lneq g
	\nonumber
\end{align} 
In this regime and within leading order in $g$, the dynamics stays close to a centre manifold approximately corresponding to
(\ref{noact:c1}).  The manifold is a centre manifold: it has an equal number of exponentially stable and unstable directions along which it can be reached or left by extremals in order to satisfy the boundary conditions. The universal properties we set out to exhibit occur when time evolution mostly occurs close to the centre manifold. Overall, for $g$ tending to zero the dynamics becomes slow-fast,
swiftly connecting the centre manifold either to the boundary conditions or to the penalty-dependant regimes qualitatively similar to those produced by (\ref{push:bv}).

\subsection{Dynamics in Fenichel's coordinates}

The first step is to introduce Fenichel's coordinates \cite{FenN1979} (see \cite{GuKu2009,CaTe2017} for recent reviews) in order to simply characterise the centre manifold. Based on the analysis of section~\ref{sec:noact}, we express the extremals in Fenichel's coordinates by setting 
\begin{align}
	&	\mathscr{f}_{t}^{\smi{1}}=\mathscr{y}_{t}^{\smi{4}}
	\nonumber\\
	&	\mathscr{f}_{t}^{\smi{2}}=\mathscr{y}_{t}^{\smi{2}}\,\mathscr{x}_{t}^{\smi{1}} +2 \,\mathscr{y}_{t}^{\smi{3}}\, \mathscr{x}_{t}^{\smi{2}} 
	\nonumber
\end{align}
and
\begin{align}
	\mathscr{f}_{t}^{\smi{3}}&= \mathscr{x}_{t}^{\smi{2}}+\varepsilon\,\mathscr{y}_{t}^{\smi{3}}\,\mathscr{x}_{t}^{\smi{3}}-\varepsilon\,\mathscr{y}_{t}^{\smi{1}}\,\mathscr{x}_{t}^{\smi{1}}
	\nonumber\\
	\mathscr{f}_{t}^{\smi{4}}&=\mathscr{x}_{t}^{\smi{1}} \left(\mathscr{x}_{t}^{\smi{2}}-2 \,\varepsilon \, \left( \mathscr{x}_{t}^{\smi{3}}+\mathscr{y}_{t}^{\smi{3}}\right)\right)+\varepsilon \,\mathscr{x}_{t}^{\smi{4}} \mathscr{x}_{t}^{\smi{1}}{}^2+2\, \varepsilon  \,
	\mathscr{x}_{t}^{\smi{2}}{}^2
	\nonumber
\end{align}
and solving for the Lagrange multipliers. In doing so, we take advantage of the fact that the Lagrange multipliers obey linear dynamics by construction. After straightforward calculations we arrive at the set of equations:
\begin{align}
	\begin{split}
			\dot{\mathscr{f}}_{t}^{\smi{1}}&=\varepsilon\,  \mathscr{f}_{t}^{\smi{2}}
		\\
		\dot{\mathscr{f}}_{t}^{\smi{2}}&=\mathscr{f}_{t}^{\smi{2}}+2\, \mathscr{f}_{t}^{\smi{3}}
		\\
		\dot{\mathscr{f}}_{t}^{\smi{3}}&=-\frac{\varepsilon\,  \left(\varepsilon\,  \mathscr{f}_{t}^{\smi{2}} \left(\mathscr{x}_{t}^{\smi{3}}+\mathscr{x}_{t}^{\smi{1}} \mathscr{x}_{t}^{\smi{4}}\right)-2 \,\mathscr{f}_{t}^{\smi{3}}
			\mathscr{x}_{t}^{\smi{2}}\right)+\mathscr{f}_{t}^{\smi{4}}}{\mathscr{x}_{t}^{\smi{1}}}
		\\
		\dot{\mathscr{f}}_{t}^{\smi{4}}&=\frac{\varepsilon\,  \mathscr{f}_{t}^{\smi{1}} \mathscr{x}_{t}^{\smi{1}}{}^2}{g}+\varepsilon\,  \delta_{t}\,\mathscr{x}_{t}^{\smi{1}}{}^2+\mathscr{f}_{t}^{\smi{4}} \left(\frac{4 \,\varepsilon\,  \mathscr{x}_{t}^{\smi{2}}}{\mathscr{x}_{t}^{\smi{1}}}+2\right)+2 \,\varepsilon^2 \,\mathscr{f}_{t}^{\smi{2}}
		\\
		&
		+2\,\varepsilon\, \mathscr{x}_{t}^{\smi{2}}\left(4\,\varepsilon\,\mathscr{x}_{t}^{\smi{3}} -5\,\mathscr{x}_{t}^{\smi{2}}\right)
		+\mathscr{x}_{t}^{\smi{1}}\left(9\,\varepsilon\,\mathscr{x}_{t}^{\smi{3}}-3\,\mathscr{x}_{t}^{\smi{2}}-6\,\varepsilon\right)-3\, \varepsilon\,  \mathscr{x}_{t}^{\smi{1}}{}^2 \,\mathscr{x}_{t}^{\smi{4}}
		-\frac{8\,
			\varepsilon^2 \,\mathscr{x}_{t}^{\smi{2}}{}^3}{\mathscr{x}_{t}^{\smi{1}}}
	\end{split}
	\label{centre:f}
\end{align}
and
\begin{align}
	\begin{split}
			\dot{\mathscr{x}}_{t}^{\smi{1}}&=2 \,\varepsilon\,  \mathscr{x}_{t}^{\smi{2}}
		\\
		\dot{\mathscr{x}}_{t}^{\smi{2}}&=\varepsilon\,  \left(\mathscr{x}_{t}^{\smi{3}}-\mathscr{x}_{t}^{\smi{1}} \mathscr{x}_{t}^{\smi{4}}\right)-\mathscr{x}_{t}^{\smi{2}}
		\\
		\dot{\mathscr{x}}_{t}^{\smi{3}}&=-2 \left(\varepsilon\,  \mathscr{x}_{t}^{\smi{2}}	\mathscr{x}_{t}^{\smi{4}}+\mathscr{x}_{t}^{\smi{3}}-1\right)
		\\
		\dot{\mathscr{x}}_{t}^{\smi{4}}&=	\frac{\mathscr{f}_{t}^{\smi{1}}}{g} +\delta_{t}
	\end{split}
	\label{centre:x}
\end{align}
In (\ref{centre:f}), (\ref{centre:x})  we define
\begin{align}
	\delta_{t}=b(\mathscr{f}_{t}^{\smi{1}})-\frac{\mathscr{f}_{t}^{\smi{1}}}{g}
	\nonumber
\end{align}
This quantity is responsible for non-linear corrections to the dynamics in the complement subspace to the centre manifold. For this reason, in what follows, we focus only on the harmonic penalty \eqref{penalty:harmonic} with $\Lambda=\sqrt{2}$ such that $\delta_t=0$. The aim is to reduce (\ref{centre:f}), (\ref{centre:x}) to a simpler slow-fast system of equations that, while 
still satisfying arbitrary boundary conditions, captures universal traits of the dynamics independently of the penalty imposed on controls.
The reader not interested in technical details of the analysis may want to directly proceed to section~\ref{sec:slow-fast}.  

\subsection{Scaling analysis}

The system of differential equations formed by (\ref{centre:f}), (\ref{centre:x}) is singular as $g$ tends to zero.
In order to study the limit, we need to find a proper rescaling of time and dynamical variables which allow us to expand the dynamics around $g$ equal zero. To this end, insight from \cite[Suppl]{MGSc2014} suggests to identify 
\begin{align}
	h=g^{1/4}
	\nonumber
\end{align} 
as the order parameter of the expansion, and describe the evolution in terms of
\begin{align}
	&	\begin{split}
		&	\mathscr{f}_{t}^{\smi{i}}=h^{5-i}\phi_{\mathscr{t}}^{\smi{i}}
		\\
		& \mathscr{x}_{t}^{\smi{i}}=\xi_{\mathscr{t}}^{\smi{i}}
	\end{split}
	&&i=1,\dots,4
	\nonumber
\end{align}
depending on  the {\textquotedblleft} fast {\textquotedblright} time variable
\begin{align}
	\mathscr{t}=\frac{t}{h}
	\nonumber
\end{align}
The starting point of the expansion becomes the system of equations
\begin{align}
	&	\frac{\mathrm{d}\phi_{\mathscr{t}}^{\smi{1}}}{\mathrm{d}\mathscr{t}}
	=\varepsilon\,  \phi_{\mathscr{t}}^{\smi{2}}
	\nonumber\\
	&	\frac{\mathrm{d}\phi_{\mathscr{t}}^{\smi{2}}}{\mathrm{d}\mathscr{t}}=h \phi_{\mathscr{t}}^{\smi{2}}+2 \phi_{\mathscr{t}}^{\smi{3}}
	\nonumber\\
	&	\frac{\mathrm{d}\phi_{\mathscr{t}}^{\smi{3}}}{\mathrm{d}\mathscr{t}}=-\frac{h^2 \varepsilon\, ^2 \left(\xi_{\mathscr{t}}^{\smi{3}}+\xi_{\mathscr{t}}^{\smi{1}} \xi_{\mathscr{t}}^{\smi{4}}\right) \phi_{\mathscr{t}}^{\smi{2}}}{\xi_{\mathscr{t}}^{\smi{1}}}+\frac{2 h \varepsilon\, 
		\xi_{\mathscr{t}}^{\smi{2}} \phi_{\mathscr{t}}^{\smi{3}}}{\xi_{\mathscr{t}}^{\smi{1}}}-\frac{\phi_{\mathscr{t}}^{\smi{4}}}{\xi_{\mathscr{t}}^{\smi{1}}}
	\nonumber\\
	&	\frac{\mathrm{d}\phi_{\mathscr{t}}^{\smi{4}}}{\mathrm{d}\mathscr{t}}=
	\varepsilon\,  \xi_{\mathscr{t}}^{\smi{1}}{}^2 \,\phi_{\mathscr{t}}^{\smi{1}}+
	2\, h^3\, \varepsilon\, ^2 \phi_{\mathscr{t}}^{\smi{2}}+2 \,h \,\phi_{\mathscr{t}}^{\smi{4}} \left(\frac{2 \varepsilon\,  \xi_{\mathscr{t}}^{\smi{2}}}{\xi_{\mathscr{t}}^{\smi{1}}}+1\right)
	\nonumber\\
	&
	+2\,\varepsilon\,\xi_{\mathscr{t}}^{\smi{2}}\left(4\,\varepsilon\,\xi_{\mathscr{t}}^{\smi{3}}-5\,\xi_{\mathscr{t}}^{\smi{2}}\right)+\xi_{\mathscr{t}}^{\smi{1}} \left(9 \varepsilon\,  \xi_{\mathscr{t}}^{\smi{3}}-3
	\xi_{\mathscr{t}}^{\smi{2}}-6 \varepsilon\, \right)-3\,\varepsilon\,  \xi_{\mathscr{t}}^{\smi{1}}{}^2\, \xi_{\mathscr{t}}^{\smi{4}}-\frac{8\,
		\varepsilon\, ^2 \xi_{\mathscr{t}}^{\smi{2}}{}^3}{\xi_{\mathscr{t}}^{\smi{1}}}
	\nonumber
\end{align}
and
\begin{align}
	&	\frac{\mathrm{d}\xi_{\mathscr{t}}^{\smi{1}}}{\mathrm{d}\mathscr{t}}=2\, h \,\varepsilon\,  \xi_{\mathscr{t}}^{\smi{2}}
	\nonumber\\
	&	\frac{\mathrm{d}\xi_{\mathscr{t}}^{\smi{2}}}{\mathrm{d}\mathscr{t}}=h \left(\varepsilon\,  \big{(}\xi_{\mathscr{t}}^{\smi{3}}-\xi_{\mathscr{t}}^{\smi{1}} \xi_{\mathscr{t}}^{\smi{4}}\big{)}-\xi_{\mathscr{t}}^{\smi{2}}\right)
	\nonumber\\
	&\frac{\mathrm{d}\xi_{\mathscr{t}}^{\smi{3}}}{\mathrm{d}\mathscr{t}}=-\,2 \,h\, \left(\varepsilon\,  \xi_{\mathscr{t}}^{\smi{2}}
	\xi_{\mathscr{t}}^{\smi{4}}+\xi_{\mathscr{t}}^{\smi{3}}-1\right)
	\nonumber\\
	&	\frac{\mathrm{d}\xi_{\mathscr{t}}^{\smi{4}}}{\mathrm{d}\mathscr{t}}=h\, \phi_{\mathscr{t}}^{\smi{1}}
	\nonumber
\end{align}
We look for a solution of the above system of differential equations in the form
\begin{align}
	&
	\begin{split}
		&	\phi_{\mathscr{t}}^{\smi{i}}=\phi_{\mathscr{t},\timt}^{\smp{i}{0}}+h\,\phi_{\mathscr{t},\timt}^{\smp{i}{0}}+O(h^{2})
		\nonumber\\
		&\xi_{\mathscr{t}}^{\smi{i}}=\xi_{\mathscr{t},\timt}^{\smp{i}{0}}+h\,\xi_{\mathscr{t},\timt}^{\smp{i}{0}}+O(h^{2})
	\end{split}
	&& i=1,\dots,4
	\nonumber
\end{align}
The time derivative with respect to $\mathscr{t}$ acts on the expansion as
\begin{align}
	\frac{\mathrm{d}}{\mathrm{d} \mathscr{t}}=\partial_{\mathscr{t}}+h \,\partial_{t}
	\nonumber
\end{align}
According to multiscale perturbation theory \cite{VerF2005}, we determine the dependence upon the {\textquotedblleft}slow time{\textquotedblright} $\timt$ by subtracting resonant terms from the regular expansion in powers of $h$.

\subsection{Zero order solution}
 \label{sec:zero}
 
 At zero order, we get
 	\begin{align}
 	&	\partial_{\mathscr{t}}	\xi_{\mathscr{t},\timt}^{\smp{i}{0}}=0 && i=1,\dots,4
 	\nonumber
 \end{align}
 which show that the expansion of the cumulants starts with pure functions of the {\textquotedblleft}slow{\textquotedblright} time $\timt$:  
 \begin{align}
 &	\xi_{\mathscr{t},\timt}^{\smp{i}{0}}=X_{\timt}^{\smi{i}} && \mathfrak{i}=\mathfrak{1},\dots,\mathfrak{4}
 	\nonumber
 \end{align}
 As expected from our qualitative analysis, only Fenichel's coordinates depend on the {\textquotedblleft}fast{\textquotedblright} time $\mathscr{t}$ and satisfy the linear non-homogeneous equation 
	\begin{align}
		\partial_{\mathscr{t}}\begin{bmatrix}
			\phi_{\mathscr{t},\timt}^{\smp{1}{0}}	\\  \phi_{\mathscr{t},\timt}^{\smp{2}{0}}	\\  \phi_{\mathscr{t},\timt}^{\smp{3}{0}}	\\  \phi_{\mathscr{t},\timt}^{\smp{4}{0}}	
		\end{bmatrix}
		=\mathsf{A}^{\smi{0}}
		\begin{bmatrix}
			\phi_{\mathscr{t},\timt}^{\smp{1}{0}}	\\  \phi_{\mathscr{t},\timt}^{\smp{2}{0}}	\\  \phi_{\mathscr{t},\timt}^{\smp{3}{0}}	\\  \phi_{\mathscr{t},\timt}^{\smp{4}{0}}	
		\end{bmatrix}
		+
		\begin{bmatrix}
			0	\\  0	\\  0	\\  F_{\timt}^{\smp{0}{0}}
		\end{bmatrix}
		\label{center:zero}
	\end{align}
	where the $4\,\times\,4$ real matrix
	\begin{align}
		\mathsf{A}^{\smi{0}}=	\begin{bmatrix}
			0 & \varepsilon & 0 & 0	\\  0 & 0 & 2 & 0 \\ 0 & 0 & 0 &  -\,\frac{1}{X_{\timt}^{\smi{1}}} \\
			\varepsilon\,  X_{\timt}^{\smi{1}}{}^2 & 0 & 0 & 0
		\end{bmatrix}
		\nonumber
	\end{align}
	and  the vanishing non-homogeneous term
	\begin{align}
		F_{\timt}^{\smp{0}{0}}&=2\,\varepsilon\,X_{\timt}^{\smi{2}}\,\left(4\,\varepsilon\,X_{\timt}^{\smi{3}}-5\,X_{\timt}^{\smi{2}}\right)
		\nonumber\\
&		+X_{\timt}^{\smi{1}} \left(9\, \varepsilon\,  X_{\timt}^{\smi{3}}-3\,
X_{\timt}^{\smi{2}}-6\, \varepsilon\, \right)
-3 \,\varepsilon\,  X_{\timt}^{\smi{1}}{}^2 \,X_{\timt}^{\smi{4}}	-\frac{8\,
			\varepsilon\, ^2 X_{\timt}^{\smi{2}}{}^3}{X_{\timt}^{\smi{1}}}
		\label{center:nh}
	\end{align}
	are constant with respect to the fast time $\mathscr{t}$. The system of differential equations (\ref{center:zero}) is thus time-autonomous and we can integrate it by diagonalizing $\mathsf{A}^{\smi{0}}$. The roots of the fourth order characteristic polynomial 	occur in pairs of complex conjugate eigenvalues that fortunately  admit an elementary expression
	\begin{align}
		\operatorname{Sp}\mathsf{A}^{0}=a\,\left\{ - (1+\imath)\,,  -(1-\imath), 1-\imath\,, 1+\imath\right\}
		\nonumber
	\end{align}
	with
	\begin{align}
		a= \left (\frac{\varepsilon^{2}\,X_{\timt}^{\smi{1}}}{2}\right )^{1/4}
		\nonumber
	\end{align}
	corresponding to two exponentially stable and unstable focuses. We define
	\begin{align}
		\begin{split}
					&		\phi_{\mathscr{t}}^{\smi{1}}=e^{-a\,(1+\imath)\,\mathscr{t}}\left(\Phi_{0,\timt}^{\smi{1}}-\frac{F_{\timt}^{\smp{0}{0}}}{a\,(1+\imath)}\right)+\frac{F_{\timt}^{\smp{0}{0}}}{a\,(1+\imath)}
		\\	&\phi_{\mathscr{t}}^{\smi{2}}=e^{a\,(1-\imath)\,\mathscr{t}}\left(\Phi_{0,\timt}^{\smi{2}}+\frac{F_{\timt}^{\smp{0}{0}}}{a\,(1-\imath)}\right)-\frac{F_{\timt}^{\smp{0}{0}}}{a\,(1-\imath)}
		\end{split}
	\label{center:zerosol}
	\end{align}
	and, after standard calculations, we find
	\begin{align}
		\phi_{\mathscr{t},\timt}^{\smp{1}{0}}&=2^{3/4}\frac{\operatorname{Im}\left(\Phi_{\timt}^{\smi{1}}+\Phi_{\timt}^{\smi{2}}\right)-\operatorname{Re}\left(\Phi_{\timt}^{\smi{1}}-\Phi_{\timt}^{\smi{2}}\right)
			}{\varepsilon^{1/2} X_{\timt}^{\smi{1}}{}^{7/4}} 
		\nonumber\\
		\phi_{\mathscr{t},\timt}^{\smp{2}{0}}&
			=\frac{2^{3/2}}{\varepsilon  X_{\timt}^{\smi{1}}{}^{3/2}}\operatorname{Im}
		\left(\phi_{\mathscr{t}}^{\smi{2}}-\phi_{\mathscr{t}}^{\smi{1}}\right)
		\nonumber\\
		\phi_{\mathscr{t},\timt}^{\smp{3}{0}}&=-\frac{2^{1/4}\,\operatorname{Im}\left(\Phi_{\timt}^{\smi{1}}+\Phi_{\timt}^{\smi{2}}\right)}{\varepsilon^{1/2} X_{\timt}^{\smi{1}}{}^{5/4}\,\imath}
		\nonumber\\
		\phi_{\mathscr{t},\timt}^{\smp{4}{0}}&
		=2 \,\operatorname{Re}\left(\phi_{\mathscr{t}}^{\smi{1}}+\phi_{\mathscr{t}}^{\smi{2}}\right)
		\nonumber
	\end{align}
    Inspection of (\ref{center:zerosol}) shows that the conditions
    \begin{align}
    	\begin{split}
    		 &  	\Phi_{0,\timt}^{\smp{1}{0}}-\frac{F_{\timt}^{\smp{0}{0}}}{a\,(1+\imath)}=0
    		\\
    		&\Phi_{0,\timt}^{\smp{2}{0}}+\frac{F_{\timt}^{\smp{0}{0}}}{a\,(1-\imath)}=0
    	\end{split}
    	\label{center:sm}
    \end{align}
    specify a center (or slow) manifold with respect to the fast time $\mathscr{t}$. On the slow manifold, the solution of (\ref{center:zero}) reduces to
	\begin{align}
		\phi_{\mathscr{t},\timt}^{\smp{1}{0}}&
		=-\frac{F_{\timt}^{\smp{0}{0}}}{\varepsilon\,X_{\timt}^{\smi{1}}{}^{2}}
		\nonumber\\
		\phi_{\mathscr{t},\timt}^{\smp{2}{0}}&=
		\phi_{\mathscr{t},\timt}^{\smp{3}{0}} =
		\phi_{\mathscr{t},\timt}^{\smp{4}{0}}=0
		\nonumber
	\end{align}
	the stationary solution (fixed point) of (\ref{center:zero}). 
	
\subsection{First order solution}
	
	First order correction to cumulants and Fenichel variables are determined by the solution of a $8\,\times\,8$ system of differential equations in the fast time:
	\begin{align}
		\partial_{\mathscr{t}}\begin{bmatrix}
			\phi_{\mathscr{t},\timt}^{\smp{1}{1}}	\\  \phi_{\mathscr{t},\timt}^{\smp{2}{1}}	\\  \phi_{\mathscr{t},\timt}^{\smp{3}{1}}	\\  \phi_{\mathscr{t},\timt}^{\smp{4}{1}}	\\	\xi_{\mathscr{t},\timt}^{\smp{1}{1}}\\	\xi_{\mathscr{t},\timt}^{\smp{2}{1}}\\	\xi_{\mathscr{t},\timt}^{\smp{4}{1}}
		\end{bmatrix}
		=\mathsf{A}^{\smi{1}}
		\begin{bmatrix}
			\phi_{\mathscr{t},\timt}^{\smp{1}{1}}	\\  \phi_{\mathscr{t},\timt}^{\smp{2}{1}}	\\  \phi_{\mathscr{t},\timt}^{\smp{3}{1}}	\\  \phi_{\mathscr{t},\timt}^{\smp{4}{1}}	\\	\xi_{\mathscr{t},\timt}^{\smp{1}{1}}\\	\xi_{\mathscr{t},\timt}^{\smp{2}{1}}\\	\xi_{\mathscr{t},\timt}^{\smp{4}{1}}
		\end{bmatrix}
		+
		\begin{bmatrix}
			\dots	\\  \dots	\\  \dots	\\  \dots \\ 	2\, \varepsilon\,  X_{\timt}^{\smi{2}}-\partial_{\timt}X_{\timt}^{\smi{1}}
			\\
			\left(\varepsilon\,  \big{(}X_{\timt}^{\smi{3}}-X_{\timt}^{\smi{1}} X_{\timt}^{\smi{4}}\big{)}-X_{\timt}^{\smi{2}}\right)
			-\partial_{\timt}X_{\timt}^{\smi{2}}
			\\
			-\,2 \,\left(\varepsilon\,  X_{\timt}^{\smi{2}}
			X_{\timt}^{\smi{4}}+X_{\timt}^{\smi{3}}-1\right)
			-\partial_{\timt}X_{\timt}^{\smi{3}}
			\\
			\phi_{\mathscr{t},\timt}^{\smp{1}{0}}
			-	\partial_{\mathscr{t}}X_{\timt}^{\smi{4}}
		\end{bmatrix}
		\label{center:one}
	\end{align}
The $8\,\times\,8$ real matrix takes the block structure
\begin{align}
	\mathsf{A}^{\smi{1}}=\begin{bmatrix}
		\mathsf{A}^{\smp{1}{1,1}} 	& \mathsf{A}^{\smp{1}{1,2}} \\  \mathsf{0}_{4} &\mathsf{0}_{4} 
	\end{bmatrix}
	\label{center:A1}
\end{align}
with $\mathsf{0}_{4}$ denoting a $4\,\times\,4$ matrix with null entries. The explicit form of the square $4\,\times\,4$ blocks
$\mathsf{A}^{\smp{1}{1,1}} $, $\mathsf{A}^{\smp{1}{1,2}} $ as well as the components of the non-homogeneous term in
(\ref{center:one}) denoted by $\dots$ do not play a role in determining the universal form of the slow-fast dynamics produced by the limit $g=h^{4}$ tending to zero. 
The universal form of the slow-fast dynamics in this limit depends only on the solvability condition of the stationary form of (\ref{center:one}). The solvability condition, as usual in multi-scale perturbation theory, fixes the dynamics of the slow time \cite{PaSt2008}.

\subsubsection{Solvability condition}
We construct perturbatively the centre manifold as the stationary solution of the dynamics with respect to the fast time.
We observe that the algebraic equation 
\begin{align}
	\mathsf{A}^{\smi{1}}\bm{v}+\bm{f}=0
	\nonumber
\end{align}
admits, by Fredholm alternative theorem, a unique solution only if $\bm{f}$ is orthogonal to the kernel $\operatorname{ker}A^{\smi{1}}{}^{\top}$ of the adjoint of $ \mathsf{A}^{\smi{1}}$. In other words for any $\bm{\tilde{v}}$ satisfying
\begin{align}
	\mathsf{A}^{\smi{1}}{}^{\top}\bm{\tilde{v}}=0
	\nonumber
\end{align}
it is necessary that
\begin{align}
	\left \langle\,\bm{\tilde{v}}\,,\mathsf{A}^{\smi{1}}\bm{v}+\bm{f}\,\right\rangle=
	\left \langle\,\bm{\tilde{v}}\,,\bm{f}\,\right\rangle=0
	\nonumber
\end{align}
The kernel of the adjoint of (\ref{center:A1}) is spanned by $\left\{ \bm{e}_{i} \right\}_{i=5}^{8}$ where the $\bm{e}_{i}$'s 
are the elements of the canonical basis of $\mathbb{R}^{8}$.  Projecting the non-homogeneous term in (\ref{center:one}) on the span of $\operatorname{ker}A^{\smi{1}}{}^{\top}$ and imposing the zero order slow manifold conditions (\ref{center:sm})
yields
\begin{align}
	&\partial_{\timt}X_{\timt}^{\smi{1}}=2\,	X_{\timt}^{\smi{2}}
	\nonumber\\
	&\partial_{\timt}X_{\timt}^{\smi{2}}=	\varepsilon\,  \big{(}X_{\timt}^{\smi{3}}-X_{\timt}^{\smi{1}} X_{\timt}^{\smi{4}}\big{)}-X_{\timt}^{\smi{2}}
	\nonumber
	\\
	&\partial_{\timt}X_{\timt}^{\smi{3}}=	-\,2 \,\left(\varepsilon\,  X_{\timt}^{\smi{2}}
	X_{\timt}^{\smi{4}}+X_{\timt}^{\smi{3}}-1\right)
	\nonumber
	\\
	&\partial_{\timt}X_{\timt}^{\smi{4}}=	-\frac{F_{\timt}^{\smp{0}{0}}}{\varepsilon\,X_{\timt}^{\smi{1}}{}^{2}}
	\nonumber
\end{align}
where the right hand side of the last equation is given by (\ref{center:nh}). We thus recover the system of differential equations (\ref{noact:reqs}) stemming from the stationary condition (\ref{noact:a1}), using centre manifold analysis and multiscale pertubation theory.

\subsection{Uniform slow-fast approximation}
\label{sec:slow-fast}

 Within leading order accuracy as $g$ tends to zero we may summarise the result of the invariant (centre) manifold analysis in a uniform slow-fast approximation to the dynamics. We get
\begin{align}
\begin{split}
	&	g^{1/4}\,\partial_{t}\mathscr{F}_{t}^{\smi{1}}=\varepsilon\,  \mathscr{F}_{t}^{\smi{2}}
	\\
	&	g^{1/4}\,\partial_{t}\mathscr{F}_{t}^{\smi{2}}=2 \,\mathscr{F}_{t}^{\smi{3}}
	\\
	&	g^{1/4}\,\partial_{t}\mathscr{F}_{t}^{\smi{3}}=-\frac{\mathscr{F}_{t}^{\smi{4}}}{X_{t}^{\smi{1}}}
	\\
	&	g^{1/4}\,\partial_{t}\mathscr{F}_{t}^{\smi{4}}=
	\varepsilon\,  X_{t}^{\smi{1}}{}^2 \,\mathscr{F}_{t}^{\smi{1}}
    \end{split}
	\label{sf:nonuniv}
\end{align}
and
\begin{align}
\begin{split}
&   \partial_{1}X_{t}^{\smi{1}}=2\, \varepsilon\,  X_{t}^{\smi{2}}
	\\
&   \partial_{t}X_{t}^{\smi{2}}= \varepsilon\,  \big{(}X_{t}^{\smi{3}}-X_{t}^{\smi{1}} X_{t}^{\smi{4}}\big{)}-X_{t}^{\smi{2}}
	\\
&   \partial_{t}X_{t}^{\smi{3}}=-\,2 \, \left(\varepsilon\,  X_{t}^{\smi{2}}
	X_{t}^{\smi{4}}+X_{t}^{\smi{3}}-1\right)
	\\
&	\partial_{t}X_{t}^{\smi{4}}= \mathscr{F}_{t}^{\smi{1}}
    \\
& \hspace{0.2cm}   -\frac{
		2\,\varepsilon\,X_{t}^{\smi{2}}\left(4\,\varepsilon\,X_{t}^{\smi{3}}-5\,X_{t}^{\smi{2}}\right)+X_{t}^{\smi{1}} \left(9 \varepsilon\,  X_{t}^{\smi{3}}-3
		X_{t}^{\smi{2}}-6 \varepsilon\, \right)-3\,\varepsilon\,  X_{t}^{\smi{1}}{}^2\, X_{t}^{\smi{4}}-\frac{8\,
			\varepsilon\, ^2 X_{t}^{\smi{2}}{}^3}{X_{t}^{\smi{1}}}}{\varepsilon\,X_{t}^{\smi{1}}{}^{2}}
	\end{split}
\label{sf:univ}
\end{align}
In section (\ref{sec:num}), we numerically compare the predictions to the above {\textquotedblleft}universal normal form{\textquotedblright}
of the non-linear dynamics with the exact stationary conditions occasioned by penalty models (\ref{penalty:hard}), (\ref{penalty:log}) and (\ref{penalty:harmonic}).

\section{Overdamped expansion}
\label{sec:od}
We exemplify the overdamped expansion in the case of the mean heat release (dissipation) (\ref{td:ep}) between equilibrium
states. In other words, we consider the boundary conditions (\ref{bc:MB}), (\ref{bc:isoth}) and (\ref{bc:sc2}).
In such a setup, the overdamped expansion provides somewhat  distinct asymptotics to those of the centre manifold. In fact, we can systematically carry out  the overdamped expansion once we hypothesise that the control coupling constant in (\ref{ext:eqs}), (\ref{ext:Hvf}) scales with the order parameter $\varepsilon$
\begin{align}
	g=\frac{\tilde{h}}{\varepsilon^{2}}
	\label{od:op}
\end{align}  
and we hold $\tilde{h}$ fixed. 

We infer that the overdamped asymptotic admits the multiscale form
\begin{align}
	&
	\begin{split}
		X_{t}^{\smi{i}}=\mathscr{x}_{t,\varepsilon^{2}t}^{\smp{i}{0}}+\varepsilon\,\mathscr{x}_{t,\varepsilon^{2}t}^{\smp{i}{1}}+\varepsilon^{2}\,\mathscr{x}_{t,\varepsilon^{2}t}^{\smp{i}{1}}+O(\varepsilon^{2})
		\\
		\mathscr{y}_{t}^{\smi{i}}=\mathscr{y}_{t,\varepsilon^{2}t}^{\smp{i}{0}}+\varepsilon\,\mathscr{y}_{t,\varepsilon^{2}t}^{\smp{i}{1}}+\varepsilon^{2}\,\mathscr{y}_{t,\varepsilon^{2}t}^{\smp{i}{1}}+O(\varepsilon^{2})
	\end{split}	
	&&i=1,\dots,4
	\nonumber
\end{align}
In particular, the scaling relation (\ref{od:op}) readily imposes
\begin{align}
&	X_{t}^{\smi{i}}=\mathscr{X}_{\varepsilon^{2}t}^{\smi{i}}+O(\varepsilon^{2})
&&
i=1,\dots,4
	\nonumber
\end{align}
Again we determine the functional dependence on the {\textquotedblleft}slow{\textquotedblright} time
\begin{align}
	\tim{2}=\varepsilon^{2}\,t
	\nonumber
\end{align}
to subtract from the expansion resonant terms \cite{VerF2005}. As for the centre manifold, we focus on the case of harmonic penalty on controls. 
The main results of the overdamped analysis are the cell-equations (\ref{od:cell1}), (\ref{od:cell2}). The reader familiar with multiscale analysis may want to proceed directly to section~\ref{sec:od_summary}.

\subsection{Zero order}

The overdamped limit surmises fast thermalisation of the momentum process. Indeed, upon setting $\varepsilon$ equal to zero,
we verify that
\begin{align}
	&\mathscr{x}_{t,\tim{2}}^{\smp{1}{0}}=\mathscr{x}_{0,\tim{2}}^{\smp{1}{0}}
	&&	\mathscr{x}_{t,\tim{2}}^{\smp{2}{0}}=0
	&	\mathscr{x}_{t,\tim{2}}^{\smp{3}{0}}=1
	&&\mathscr{x}_{t,\tim{2}}^{\smp{4}{0}}=\mathscr{x}_{0,\tim{2}}^{\smp{4}{0}}
	\nonumber
\end{align}
and
\begin{align}
	&\mathscr{y}_{t,\tim{2}}^{\smp{1}{0}}=\mathscr{y}_{0,\tim{2}}^{\smp{1}{0}}
	&&	\mathscr{y}_{t,\tim{2}}^{\smp{2}{0}}=\mathscr{y}_{0,\tim{2}}^{\smp{2}{0}}\,e^{t}
	&	\mathscr{y}_{t,\tim{2}}^{\smp{3}{0}}=-\frac{1}{2}
	&& \mathscr{y}_{t,\tim{2}}^{\smp{4}{0}}=\mathscr{y}_{0,\tim{2}}^{\smp{4}{0}}
	\nonumber
\end{align}
satisfy the necessary condition for optimality.

\subsection{First order}

The expansion around the zero order solution produces the equations
\begin{align}
&	\begin{cases}
	\partial_{t}\mathscr{x}_{t,\tim{2}}^{\smp{1}{1}}=0
	\\
	\partial_{t}\mathscr{x}_{t,\tim{2}}^{\smp{2}{1}}=- \mathscr{x}_{t,\tim{2}}^{\smp{2}{1}}- \left(\mathscr{x}_{0,\tim{2}}^{\smp{4}{0}}\mathscr{x}_{0,\tim{2}}^{\smp{1}{0}} -1\right)  
	\\
	\partial_{t}\mathscr{x}_{t,\tim{2}}^{\smp{3}{1}}=-2 \,\mathscr{x}_{t,\tim{2}}^{\smp{3}{1}}
		\\
	\partial_{t}\mathscr{x}_{t,\tim{2}}^{\smp{4}{1}}=0
	\end{cases}
	&&
		&	\begin{cases}
			\partial_{t}\mathscr{y}_{t,\tim{2}}^{\smp{1}{1}}=0 
			\\
			\partial_{t}\mathscr{y}_{t,\tim{2}}^{\smp{2}{1}}=-2\,\mathscr{y}_{0,\tim{2}}^{\smp{1}{0}}+\mathscr{y}_{t,\tim{2}}^{\smp{2}{1}}-\mathscr{x}_{0,\tim{2}}^{\smp{4}{0}} 
			\\
			\partial_{t}\mathscr{y}_{t,\tim{2}}^{\smp{3}{1}}=2\,\mathscr{y}_{t,\tim{2}}^{\smp{3}{1}}
			\\
			\partial_{t}\mathscr{y}_{t,\tim{2}}^{\smp{4}{1}}= \mathscr{y}_{0,\tim{2}}^{\smp{2}{0}}\,\mathscr{x}_{0,\tim{2}}^{\smp{1}{0}}\,e^{t}
	\end{cases}
	\nonumber
\end{align}
Self consistence of the expansion requires
\begin{align}
	\partial_{t}\mathscr{y}_{t,\tim{2}}^{\smp{4}{1}}=0
	\nonumber
\end{align}
which entails
\begin{align}
	\mathscr{y}_{0,\tim{2}}^{\smp{2}{0}}=0
	\nonumber
\end{align}
We then obtain
\begin{align}
	&\mathscr{x}_{t,\tim{2}}^{\smp{1}{1}}=\mathscr{x}_{0,\tim{2}}^{\smp{1}{1}}
	&&	\mathscr{x}_{t,\tim{2}}^{\smp{2}{1}}=-(1-e^{-t})\left(\mathscr{x}_{0,\tim{2}}^{\smp{4}{0}}\mathscr{x}_{0,\tim{2}}^{\smp{1}{0}} -1\right)  
	&	\mathscr{x}_{t,\tim{2}}^{\smp{3}{1}}=0
	\nonumber
\end{align}
and
\begin{align}
	&\mathscr{y}_{t,\tim{2}}^{\smp{1}{1}}=\mathscr{y}_{0,\tim{2}}^{\smp{1}{1}}
	&&	\mathscr{y}_{t,\tim{2}}^{\smp{2}{1}}=\mathscr{y}_{0,\tim{2}}^{\smp{2}{1}}\,e^{t}-(e^{t}-1)\,\left(\mathscr{x}_{0,\tim{2}}^{\smp{4}{0}}+2\, \mathscr{y}_{0,\tim{2}}^{\smp{1}{0}}\right)
	&	\mathscr{y}_{t,\tim{2}}^{\smp{3}{1}}=\mathscr{y}_{0,\tim{2}}^{\smp{3}{1}}\,e^{2\,t}
	\nonumber
\end{align}
First order corrections to stiffness and corresponding drift are constant with  respect to the faster time scale
\begin{align}
&	\mathscr{x}_{t,\tim{2}}^{\smp{4}{1}}=\mathscr{x}_{0,\tim{2}}^{\smp{1}{1}}
&& 	\mathscr{y}_{t,\tim{2}}^{\smp{4}{1}}=\mathscr{y}_{0,\tim{2}}^{\smp{1}{1}}
	\nonumber
\end{align}
We enforce equilibrium boundary conditions by requiring
\begin{align}
	\mathscr{x}_{0,0}^{\smp{4}{0}}\mathscr{x}_{0,0}^{\smp{1}{0}} = \mathscr{x}_{0,\varepsilon^{2}\,\tf}^{\smp{4}{0}}\mathscr{x}_{0,\varepsilon^{2}\,\tf}^{\smp{1}{0}} =1
	\nonumber
\end{align}
which also ensure the vanishing of the position momentum cross correlation at the end of the control horizon.

\subsection{Second order: cell equations}

The perturbative expansion produces resonances at second order. The differential equations in the slow time $\tim{2}$, obtained by subtracting secular terms arising from resonances, recover the overdamped dynamics. In the multiscale literature (see e.g. \cite{PaSt2008}), these equations are usually referred to as {\textquotedblleft}cell equations{\textquotedblright}.
Specifically, the regular expansion in $\varepsilon$ yields
\begin{align}
		\begin{cases}
		\partial_{t}\mathscr{x}_{t,\tim{2}}^{\smp{1}{2}}+\partial_{\tim{2}}\mathscr{x}_{0,\tim{2}}^{\smp{1}{0}}= -2 \,(1-e^{-t})\,\left(\mathscr{x}_{0,\tim{2}}^{\smp{4}{0}}\mathscr{x}_{0,\tim{2}}^{\smp{1}{0}} -1\right)   
		\\
		\partial_{t}\mathscr{x}_{t,\tim{2}}^{\smp{2}{2}}=- \mathscr{x}_{t,\tim{2}}^{\smp{2}{2}}-\left(
		\mathscr{x}_{0,\tim{2}}^{\smp{4}{0}}\mathscr{x}_{0,\tim{2}}^{\smp{1}{1}}
		+\mathscr{x}_{0,\tim{2}}^{\smp{4}{1}}\mathscr{x}_{0,\tim{2}}^{\smp{1}{0}}\right)
		\\
		\partial_{t}\mathscr{x}_{t,\tim{2}}^{\smp{3}{2}}=-2 \left(\mathscr{x}_{t,\tim{2}}^{\smp{3}{2}}+\mathscr{x}_{\tim{2}}^{\smp{4}{0}} (1-e^{-t})(\mathscr{x}_{\tim{2}}^{\smp{4}{0}}\mathscr{x}_{\tim{2}}^{\smp{1}{0}} -1)\right)		
				\\
		\partial_{t}\mathscr{x}_{t,\tim{2}}^{\smp{4}{2}}+\partial_{\tim{2}}\mathscr{x}_{0,\tim{2}}^{\smp{4}{0}}=\dfrac{\mathscr{y}_{0,\tim{2}}^{\smp{4}{0}}}{\tilde{h}}
	\end{cases}
	\nonumber
\end{align} 
Resonances correspond to solutions of regular perturbation theory linearly growing in the fast time variable $t$. Their presence
leads to a break down of the expansion as $t$ becomes order $O(\varepsilon^{-1})$. In the limit of infinite separation between the fast and slow time scale, we cancel such terms by requiring 
\begin{align}
	\begin{split}
&	\partial_{\tim{2}}\mathscr{x}_{0,\tim{2}}^{\smp{1}{0}}=-2 \,\left(\mathscr{x}_{0,\tim{2}}^{\smp{4}{0}}\mathscr{x}_{0,\tim{2}}^{\smp{1}{0}} -1\right)   
\\
&\partial_{\tim{2}}\mathscr{x}_{0,\tim{2}}^{\smp{4}{0}}=\dfrac{\mathscr{y}_{0,\tim{2}}^{\smp{4}{0}}}{\tilde{h}}		
	\end{split}
\label{od:cell1}
\end{align}
Furthermore by setting
\begin{align}
	\mathscr{x}_{0,\tim{2}}^{\smp{4}{1}}=\mathscr{x}_{0,\tim{2}}^{\smp{1}{1}}=0
	\nonumber
\end{align}
we are in the position to derive the prediction
\begin{align}
&\mathscr{x}_{t,\tim{2}}^{\smp{2}{1}}=-(1-e^{-t})\left(\mathscr{x}_{0,\tim{2}}^{\smp{4}{0}}\mathscr{x}_{0,\tim{2}}^{\smp{1}{0}} -1\right)  
\nonumber\\	
&	\mathscr{x}_{t,\tim{2}}^{\smp{3}{2}}=\mathscr{x}_{\tim{2}}^{\smp{4}{0}} (\mathscr{x}_{\tim{2}}^{\smp{4}{0}}\mathscr{x}_{\tim{2}}^{\smp{1}{0}} -1)
	\nonumber
\end{align}
describing within accuracy the evolution of cumulants involving the momentum process. In order to fully specify the solution, we analyze the equations satisfied by the Lagrange multipliers
\begin{align}
	\begin{cases}
		\partial_{t}\mathscr{y}_{t,\tim{2}}^{\smp{1}{2}}+\partial_{\tim{2}}\mathscr{y}_{t,\tim{2}}^{\smp{1}{0}} =\mathscr{y}_{t,\tim{2}}^{\smp{2}{1}} \,\mathscr{x}_{0,\tim{2}}^{\smp{4}{0}}
		\\
		\partial_{t}\mathscr{y}_{t,\tim{2}}^{\smp{2}{2}}=-2\,\mathscr{y}_{t,\tim{2}}^{\smp{1}{1}}+\mathscr{y}_{t,\tim{2}}^{\smp{2}{2}}+2\,\mathscr{y}_{t,\tim{2}}^{\smp{3}{1}}  \,\mathscr{x}_{0,\tim{2}}^{\smp{4}{0}}
		\\
		\partial_{t}\mathscr{y}_{t,\tim{2}}^{\smp{3}{2}}+\partial_{\tim{2}}\mathscr{y}_{t,\tim{2}}^{\smp{3}{0}}=- \mathscr{y}_{t,\tim{2}}^{\smp{2}{1}}+2\,\mathscr{y}_{t,\tim{2}}^{\smp{3}{2}}
		\\
		\partial_{t}\mathscr{y}_{t,\tim{2}}^{\smp{4}{2}}+\partial_{\tim{2}}\mathscr{y}_{0,\tim{2}}^{\smp{4}{0}} =\mathscr{y}_{t,\tim{2}}^{\smp{2}{1}}\,\mathscr{x}_{0,\tim{2}}^{\smp{1}{0}}-\mathscr{x}_{t,\tim{2}}^{\smp{2}{1}}
	\end{cases}
	\nonumber
\end{align}
Self consistency of the expansion requires
\begin{align}
	\partial_{t}\mathscr{y}_{t,\tim{2}}^{\smp{4}{2}}=0
	\nonumber
\end{align}
whence it follows
\begin{align}
	\partial_{\tim{2}}\mathscr{y}_{0,\tim{2}}^{\smp{4}{0}} =\mathscr{y}_{t,\tim{2}}^{\smp{2}{1}}\,\mathscr{x}_{0,\tim{2}}^{\smp{1}{0}}-\mathscr{x}_{t,\tim{2}}^{\smp{2}{1}}
	\nonumber
\end{align}
The right hand side must be a pure function of the slow time $\tim{2}$. This latter condition is fulfilled by setting
\begin{align}
&		\mathscr{y}_{0,\tim{2}}^{\smp{2}{1}}=\mathscr{x}_{0,\tim{2}}^{\smp{4}{0}}+2\,	\mathscr{y}_{0,\tim{2}}^{\smp{1}{0}}
\nonumber	\\
&	\mathscr{x}_{0,\tim{2}}^{\smp{2}{1}}=-\,\left(\mathscr{x}_{0,\tim{2}}^{\smp{4}{0}}\mathscr{x}_{0,\tim{2}}^{\smp{1}{0}}-1\right )
	\nonumber
\end{align}
We thus obtain the cell problem of the adjoint dynamics
\begin{align}
	\begin{split}
&	\partial_{\tim{2}}\mathscr{y}_{0,\tim{2}}^{\smp{4}{0}}=2\,\mathscr{x}_{0,\tim{2}}^{\smp{1}{0}}\,\mathscr{x}_{0,\tim{2}}^{\smp{4}{0}}+2\,\mathscr{y}_{0,\tim{2}}^{\smp{1}{0}}\,\mathscr{x}_{0,\tim{2}}^{\smp{1}{0}}-1
	\\
&\partial_{\tim{2}}\mathscr{y}_{0,\tim{2}}^{\smp{1}{0}}=\left(\mathscr{x}_{0,\tim{2}}^{\smp{4}{0}}+2\,\mathscr{y}_{0,\tim{2}}^{\smp{1}{0}}\right)\mathscr{x}_{0,\tim{2}}^{\smp{4}{0}}	
	\end{split}
	\label{od:cell2}
\end{align}

\subsection{Comparison with overdamped dynamics}
\label{sec:od_summary}
The second cumulant of an overdamped Gaussian dynamics satisfies
\begin{align}
	\dot{\mathscr{X}}_{t}^{\smi{1}}=-2\,(\mathscr{X}_{t}^{\smi{1}}\mathscr{X}_{t}^{\smi{2}}-1)
	\nonumber
\end{align}
where $ \mathscr{X}_{t}^{\smi{2}}$ is the stiffness of the mechanical potential acting on the micro-particle.
The corresponding mean dissipation functional is amenable to the form
\begin{align}
	\mathcal{E}_{\tf}=\int_{0}^{\tf}\mathrm{d}t\,\mathscr{X}_{t}^{\smi{2}}(\mathscr{X}_{t}^{\smi{1}}\mathscr{X}_{t}^{\smi{2}}-1)
	\nonumber
\end{align}
A straightforward calculation then proves that the cell equations (\ref{od:cell1}), (\ref{od:cell2}) indeed recover the overdamped optimal control conditions for an harmonic penalty on stiffness manipulations.
We emphasize that the coincidence of the equations obtained by first optimizing and then expanding with those describing the inverse order of the operations is a non-trivial result, generally not verified by applications of multiscale analysis to optimal control theory, see e.g. discussion in \cite{HaLaZhPa2014}.  Physically, the above result follows from  considering a dynamics enjoying detailed balance \cite{BoCe2017}.

\section{Uncontrolled relaxation to equilibrium}
\label{sec:relax}

If we surmise that, for times larger than the upper end $\tf$ of the control horizon, the mechanical potential is frozen at its value at $\tf$, the dynamics of second order cumulants obeys (\ref{dyn:2cum}) with $\lambda$ equal to zero, and
\begin{align}
&	\mathscr{x}_{t}^{\smi{4}}=\mathscr{x}_{\tf}^{\smi{4}} && \forall\,t\,\geq\,\tf
	\nonumber
\end{align}
The dynamics is then exactly solvable, see e.g. \cite[\S~10.2.1]{RisH1996}. Relaxation to thermal equilibrium is governed by the 
decay rates
\begin{align}
	\mathscr{r}_{\pm}=
	\begin{cases}
	\dfrac{1\pm\sqrt{1-4\,\varepsilon^{2}\,\mathscr{x}_{\tf}^{\smi{4}}}}{2}	& \mbox{if}\,\,0<\mathscr{x}_{\tf}^{\smi{4}}\,< \,\dfrac{1}{4}
		\\[0.3cm]
		\dfrac{1}{2}&\mbox{if}\,\,\mathscr{x}_{\tf}^{\smi{4}}\,\geq\, \dfrac{1}{4}
	\end{cases}
	\nonumber
\end{align}
The mean heat release during the thermalisation process is
\begin{align}
	\mathcal{Q}_{T}=\int_{\tf}^{T}\mathrm{d}t\,\left(1+\frac{\mathscr{x}_{\tf}^{\smi{4}} \, \varepsilon ^2 \left(\mathscr{x}_{\tf}^{\smi{1}}\,\mathscr{x}_{\tf}^{\smi{4}}-1\right) e^{-2\,\mathscr{r}_{-}\,(t-\tf) }\, \left(1-e^{-(\mathscr{r}_{+}-\mathscr{r}_{-})\,(t-\tf) }\right)^2}{(\mathscr{r}_{+}-\mathscr{r}_{-}) ^2}
	\right)-(T-\tf)
	\nonumber
\end{align}
The integral converges if
\begin{align}
\mathscr{x}_{\tf}^{\smi{4}}\,>\,0
	\label{relax:confining}
\end{align}
This condition is clearly always satisfied by transitions between equilibrium states. 
Physically, \eqref{relax:confining} means that once the dynamics is no longer controlled, the system is left under the action of a confining mechanical potential $U_{\tf}$. 
For $t\,\geq\,\tf$, the system will undergo a process of thermal relaxation to reach equilibrium with $U_{\tf}$.
However, if \eqref{relax:confining} does not hold and the stiffness is negative at $\tf$, then the system is left in an unstable state and its spatial probability distribution is pushed towards infinity. When the stiffness is the control, negative values may be present in theoretical optimal protocols for transitions at minimum dissipation between assigned states. Reference \cite{AlLaJu2020} describes a method to implement a potential with negative stiffness by means of optical feedback traps.

When the stiffness is positive at the end of a transition between non-equilibrium states at minimum work, i.e. (\ref{relax:confining}) holds, the mean heat release during the relaxation is
\begin{align}
	\mathcal{Q}_{\star}=\lim_{T\nearrow\infty}	\mathcal{Q}_{T,\tf}=\frac{\varepsilon ^2}{2\,\mathscr{r}_{-}\,\mathscr{r}_{+}} \,\frac{\mathscr{x}_{\tf}^{\smi{4}} \,  \left(\mathscr{x}_{\tf}^{\smi{1}}\,\mathscr{x}_{\tf}^{\smi{4}}-1\right) }{\mathscr{r}_{+}+\mathscr{r}_{-}}
	= \,\frac{\mathscr{x}_{\tf}^{\smi{1}}\,\mathscr{x}_{\tf}^{\smi{4}}-1}{2}
	\nonumber
\end{align}
which is positive as long as
\begin{align}
	\mathscr{x}_{\tf}^{\smi{1}}\,\mathscr{x}_{\tf}^{\smi{4}}\,\geq\,1
	\label{relax:contraction}
\end{align}
As no work is done on the system during the relaxation process and the temperature of the system at the end of the control horizon is in equilibrium with the environment, the first law of thermodynamics implies
\begin{align}
	0\,\leq\,\lim_{T\nearrow\infty}\mathscr{Q}_{T,\tf}= \operatorname{E}U_{\tf}(\mathscr{q}_{\tf})-\lim_{T\nearrow\infty}\operatorname{E}U_{\tf}(\mathscr{q}_{T})
	\nonumber
\end{align}
The system therefore releases heat while contracting toward a state described by a more confining probability distribution than the one reached at the end of the control horizon equilibrium. 

Independently of the value of the stiffness at the end of the control horizon, the process satisfies the second law of thermodynamics (\ref{td:ep}). Namely, upon including the Gibbs-Shannon entropy, we get
\begin{align}
\lim_{T\nearrow\infty}	\mathcal{E}_{T,\tf}=\lim_{T\nearrow\infty}\operatorname{E} U_{\tf}(\mathscr{q}_{T}) -\operatorname{E} U_{\tf}(\mathscr{q}_{\tf})+\mathcal{Q}_{\star}=\frac{1-\mathscr{x}_{\tf}^{\smi{1}}\,\mathscr{x}_{\tf}^{\smi{4}}}{2}+\frac{\mathscr{x}_{\tf}^{\smi{1}}\,\mathscr{x}_{\tf}^{\smi{4}}-1}{2}=0
	\nonumber
\end{align}

\section{Numerical Analysis}
\label{sec:num}

Numerical methods allow us to illustrate engineered swift equilibration protocols for some examples and to contrast them with protocols implementing transitions at minimum work. There are two ways in which the problem can be approached numerically: directly by minimising the cost functional, and indirectly by solving the first order optimality conditions. 
Direct methods parametrise the control: the control and state variables are discretised in time and space. Inequality constraints (Karush-Kuhn-Tucker conditions, see e.g. \cite{HaAg2014}), such as those given by \eqref{penalty:hard}, on states are enforced by logarithmic barriers. Sophisticated numerical algorithms, such as Interior Point Optimisation (IPOpt)~\cite{WacA2009,TaCoWaLa2019}, then solve a linearised system to find an optimal direction and make gradient-descent like updates to find an optimal solution.
A drawback of using direct methods is that, since these require discretisation of time and space coordinates and computation of gradients, they suffer from the curse of dimensionality, whereby increasing the number of dimensions of the problem significantly affects the computational cost. 

The indirect approach solves the necessary conditions for optimality derived from Pontryagin's maximum principle. The necessary conditions are often a two-point boundary value problem with equations for the co-state variables, such as those in \eqref{ext:eqs}. To solve these boundary value problems, we can use for example shooting methods, where the problem is integrated numerically forward from the given initial data and then trajectories are adjusted to bring them closer to the terminal conditions. Clearly, such a method is very sensitive to the initial guess. Another, generally more robust option is collocation, where the solution is approximated in a finite-dimensional set eg polynomials on a fixed set of points, called the collocation points~\cite{Ascher2015}.

\subsection{Engineered swift equilibration}
\label{sec:ep_num}

\begin{figure}[!h]  
    \centering
 \includegraphics[width=0.9\linewidth]{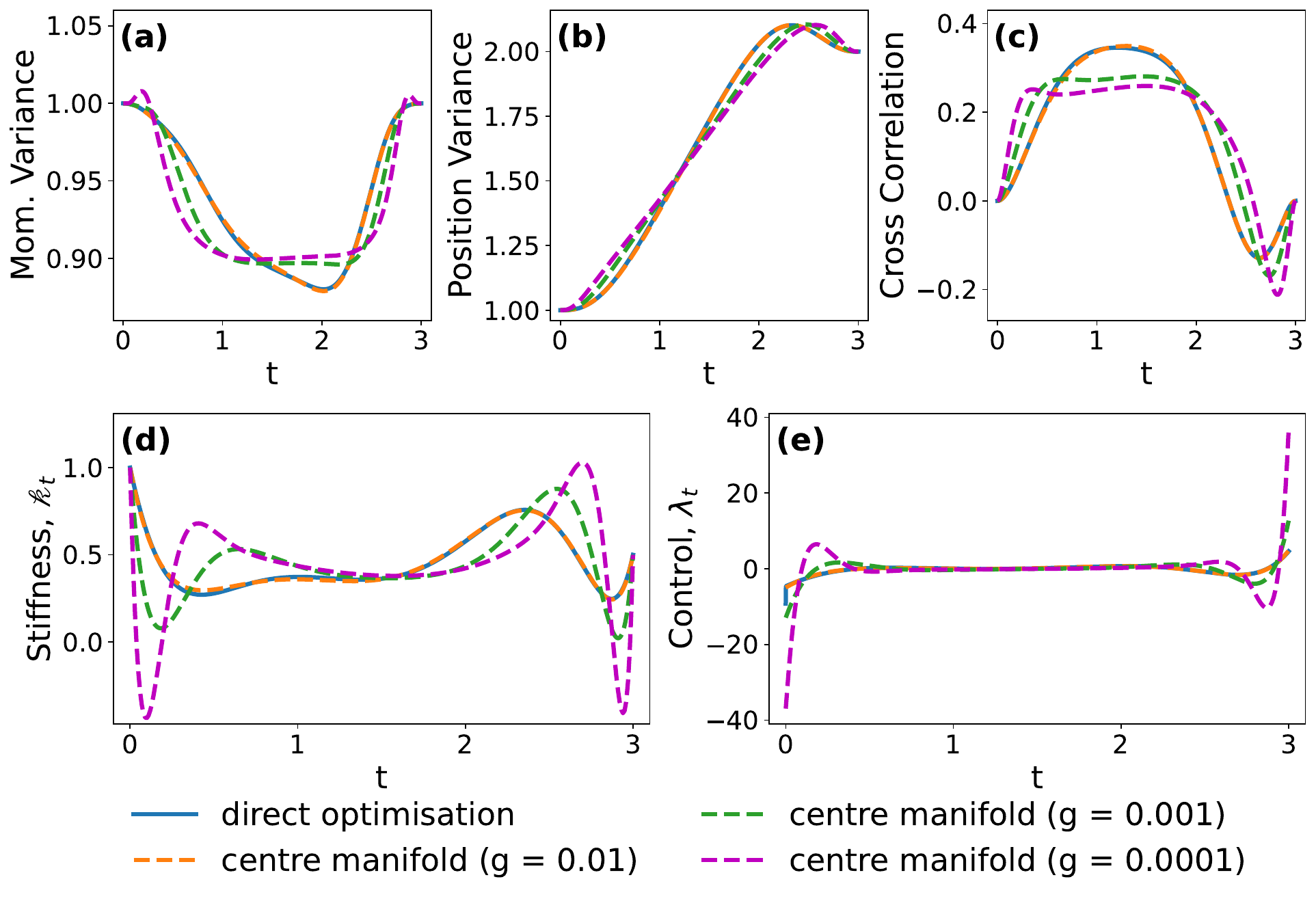}
    \caption{Engineered swift equilibration minimising the entropy production~\eqref{ep:cost}, subject to the harmonic penalty \eqref{penalty:harmonic}. The solution is computed using a direct optimisation on the cost functional~\eqref{ep:cost} (blue) with InfiniteOpt.jl~\cite{PuZhHoZa2022} and solved using Interior Point Optimisation IPOpt~\cite{WacA2009} and using $g=0.01$. The central manifold solution is shown for decreasing values of $g$: $g=0.01$ orange, $g=0.001$ green, $g=0.0001$ maroon; and is found by numerically integrating the set of differential equations~\eqref{sf:nonuniv},~\eqref{sf:univ} with a collocation method from the package DifferentialEquations.jl~\cite{rackauckas2017differentialequations,Lobatto_Jay2015}. We fix $\Lambda=\sqrt{2},\,\varepsilon=1,\,\tf=3$ and boundary conditions are described in \eqref{ep_num:bc}, with $\sigma^2_0 = 1$ and $\sigma^2_{\tf}=2$. Results produced using the code available at \cite{github_link}}
    \label{fig:harmonic}
\end{figure}

\begin{figure}[!h]    
    \centering \includegraphics[width=0.9\linewidth]{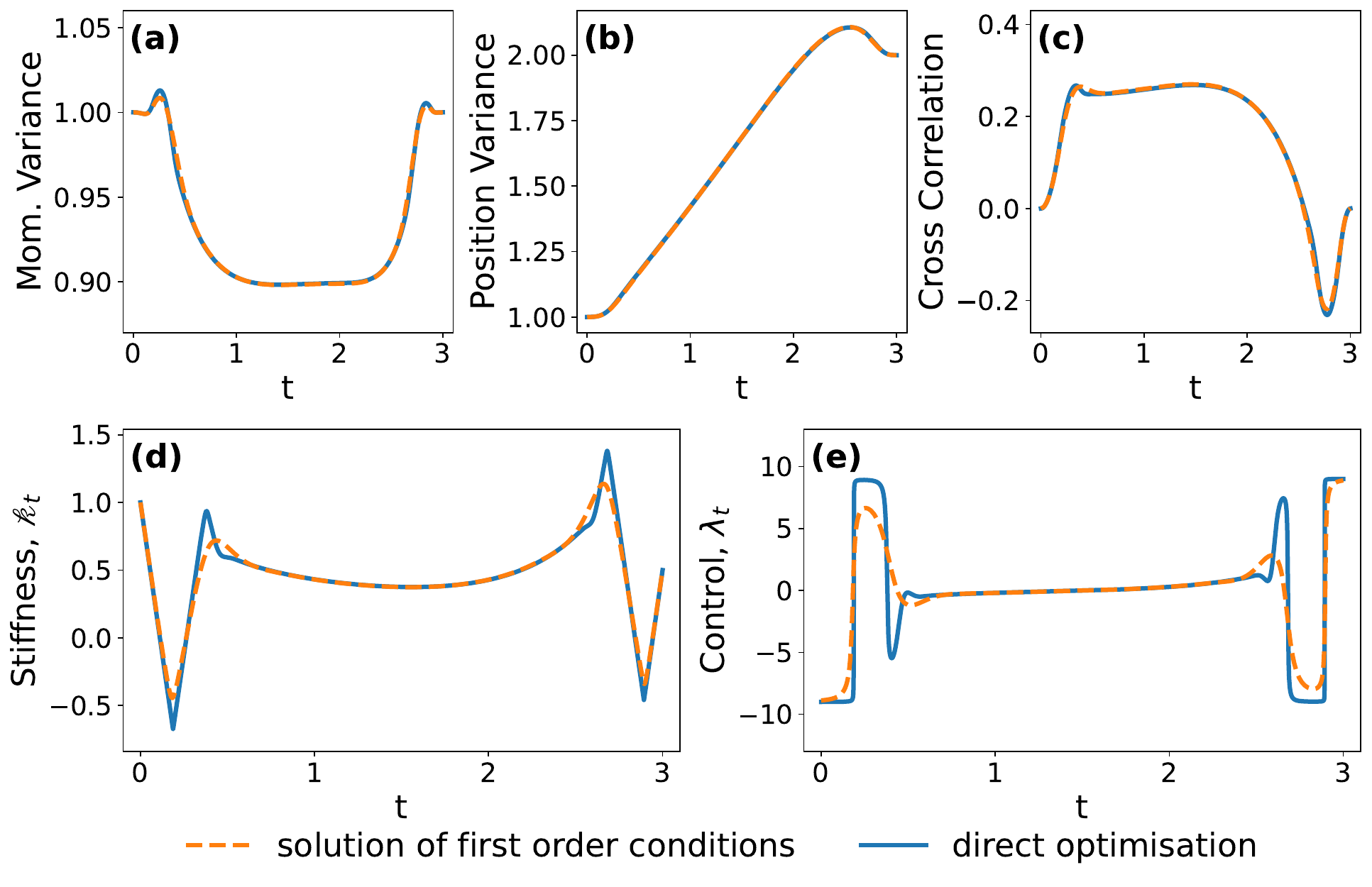}
    \caption{Engineered swift equilibration minimising the entropy production~\eqref{ep:cost}. We find the solution by a direct optimisation of the cost functional with a hard penalty~\eqref{penalty:hard} (blue) and the solution of system of differential equations specifying the first order optimality conditions~\eqref{ext:eqs} with a logarithmic penalty~\eqref{penalty:log} (orange-dashed).     
    We fix $\Lambda=9,\,\varepsilon=1,\,\tf=3$ and use $g=0.001$ and $g=0$ for the logarithmic and hard penalties respectively. The boundary conditions are given by~\eqref{noneq_bc}, with $\sigma^2_0 = 1$ and $\sigma^2_{\tf}=2$. Numerical methods are as described in Fig.~\ref{fig:harmonic}.}
    \label{fig:compactlog_method}
\end{figure}

\begin{figure}[!hb]   
    \centering \includegraphics[width=0.9\linewidth]{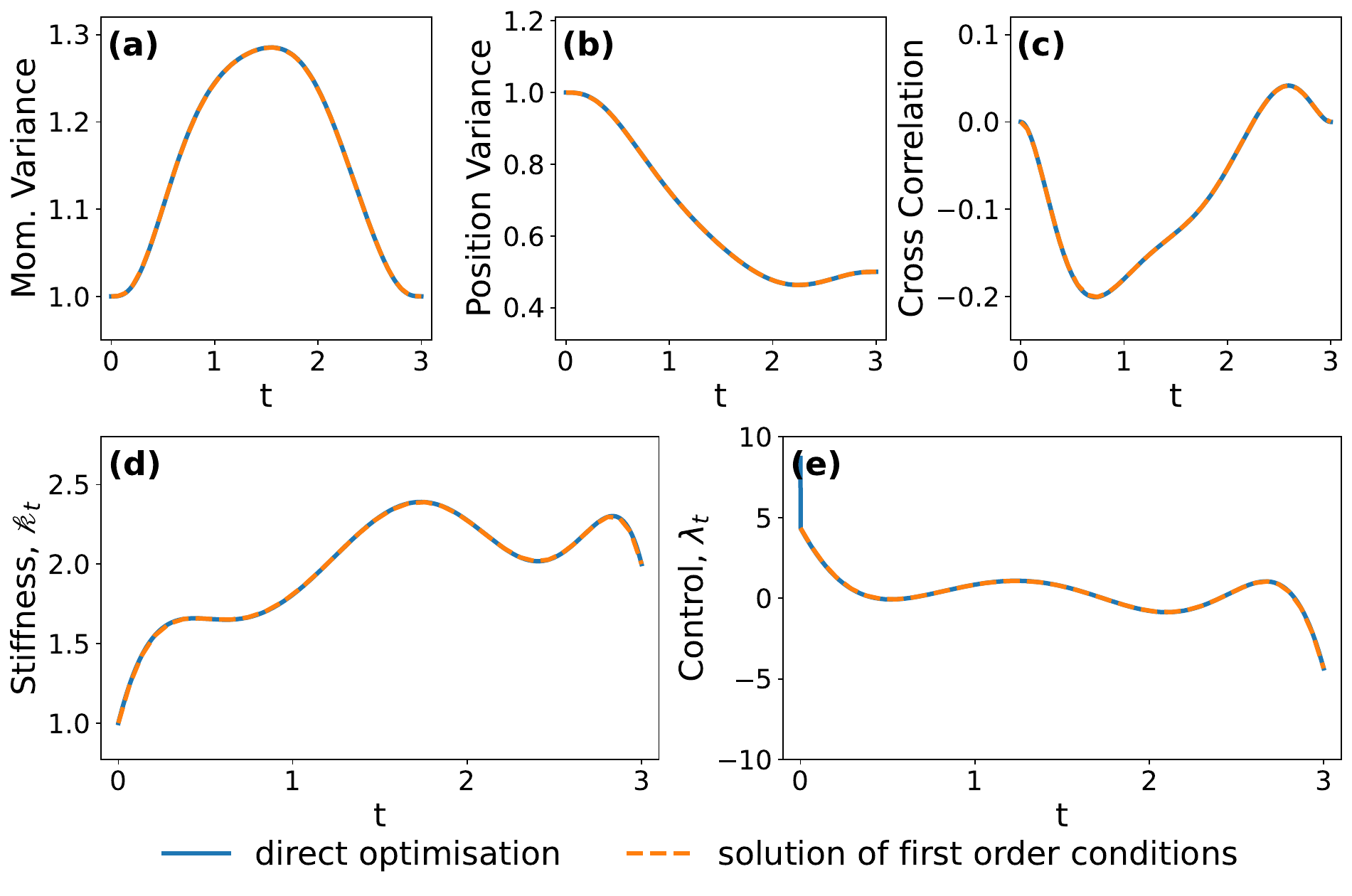}
    \caption{Engineered swift equilibration minimising the entropy production~\eqref{ep:cost} for a contraction. We use a harmonic penalty~\eqref{penalty:harmonic} and find the solution with a direct optimisation of the cost functional (blue lines) and as a solution of the first order conditions (orange, dashed).
    We fix $\Lambda=\sqrt{2},\,\varepsilon=1,\,\tf=3$ and $g=0.01$. Boundary conditions are given by~\eqref{noneq_bc}, with $\sigma^2_0 = 1$ and $\sigma^2_{\tf}=1/2$. Numerical methods are as described in Fig.~\ref{fig:harmonic}.}
    \label{fig:contract_harmonic}
\end{figure}

When the stiffness is part of the system state, a transition between two equilibrium states in the control horizon $[0,\tf]$ is defined by the boundary conditions
\begin{align}
	&	\begin{cases}
		x_{0}^{\smi{1}}=\sigma^2_0		&
		\\
		x_{0}^{\smi{2}}=0		&
		\\
		x_{0}^{\smi{3}}=1		&
		\\
		x_{0}^{\smi{4}}=1/\sigma^2_0			&
	\end{cases}
	&&
	\begin{cases}
		x_{\tf}^{\smi{1}}=\sigma^2_{\tf}&
		\\
		x_{\tf}^{\smi{2}}=0		&
		\\
		x_{\tf}^{\smi{3}}=1		&
		\\
		x_{\tf}^{\smi{4}}=1/\sigma^2_{\tf}	&
	\end{cases}
	\label{ep_num:bc}
\end{align}
We can compute the moments of the position and momentum processes, the stiffness, and control of the engineered swift equilibration using both indirect and direct numerical methods. The indirect method means finding a numerical solution to the boundary value problem given by the set of differential equations~\eqref{ext:eqs} with boundary conditions~\eqref{ep_num:bc}. The direct method minimises the cost~\eqref{ep:cost} directly while obeying the dynamical constraints and boundary conditions on the cumulants. Because in this case the cost~\eqref{ep:cost} is a strictly convex function of the control, we expect that solutions of the differential equations to the boundary conditions specifying the stationary value of the Pontryagin functional do not develop conjugate points. This observation is supported by the good agreement of different integration methods, shown in Fig.~\ref{fig:harmonic} for an expansion with the harmonic penalty~\eqref{penalty:harmonic}, in Fig.~\ref{fig:compactlog_method} with the logarithmic~\eqref{penalty:log} and hard~\eqref{penalty:hard} penalties, and for a contraction with the harmonic penalty in Fig.~\ref{fig:contract_harmonic}. Fig.~\ref{fig:harmonic} also shows the agreement of the centre manifold in Sec.~\ref{sec:slow-fast} with the direct method of optimisation. The direct method of optimisation imposes constraints via self-concordant barrier functions, similar to the logarithmic penalty given by \eqref{penalty:log}, and 
Fig.~\ref{fig:compactlog_method} demonstrates the similarities between the qualitative behaviour of the two penalties, particularly for increasing values of $\Lambda$.
 

\subsubsection{Comparison with minimum entropy production transitions when stiffness is the control}

We now contrast two situations:
\begin{enumerate}[label=\textbf{S.\Roman*}]
	\item \label{S1} Engineered swift equilibration with an additional constraint that confines the stiffness to a compact interval $[\kappa_{m},\kappa_{M}]$. 
    \item \label{S2} Minimization of entropy production directly now using the stiffness as a control subject to a hard penalty that confines it to $[\kappa_{m},\kappa_{M}]$.
\end{enumerate}
Case (\ref{S2}) is the Gaussian version of the problem studied in \cite{MGSc2014,SaBaMG2024,SaBaMG2024_2}, where we have replaced the harmonic penalty with a hard penalty. By considering the hard penalty, we can think of case (\ref{S2}) as the limit of case (\ref{S1}) when no penalty is imposed on the time differential of the stiffness. As a consequence, we expect the entropy production in case (\ref{S2}) to always provide a lower bound on the entropy production computed in (\ref{S1}). 

\begin{figure}[!ht]
    \centering
    \includegraphics[width=0.9\linewidth]{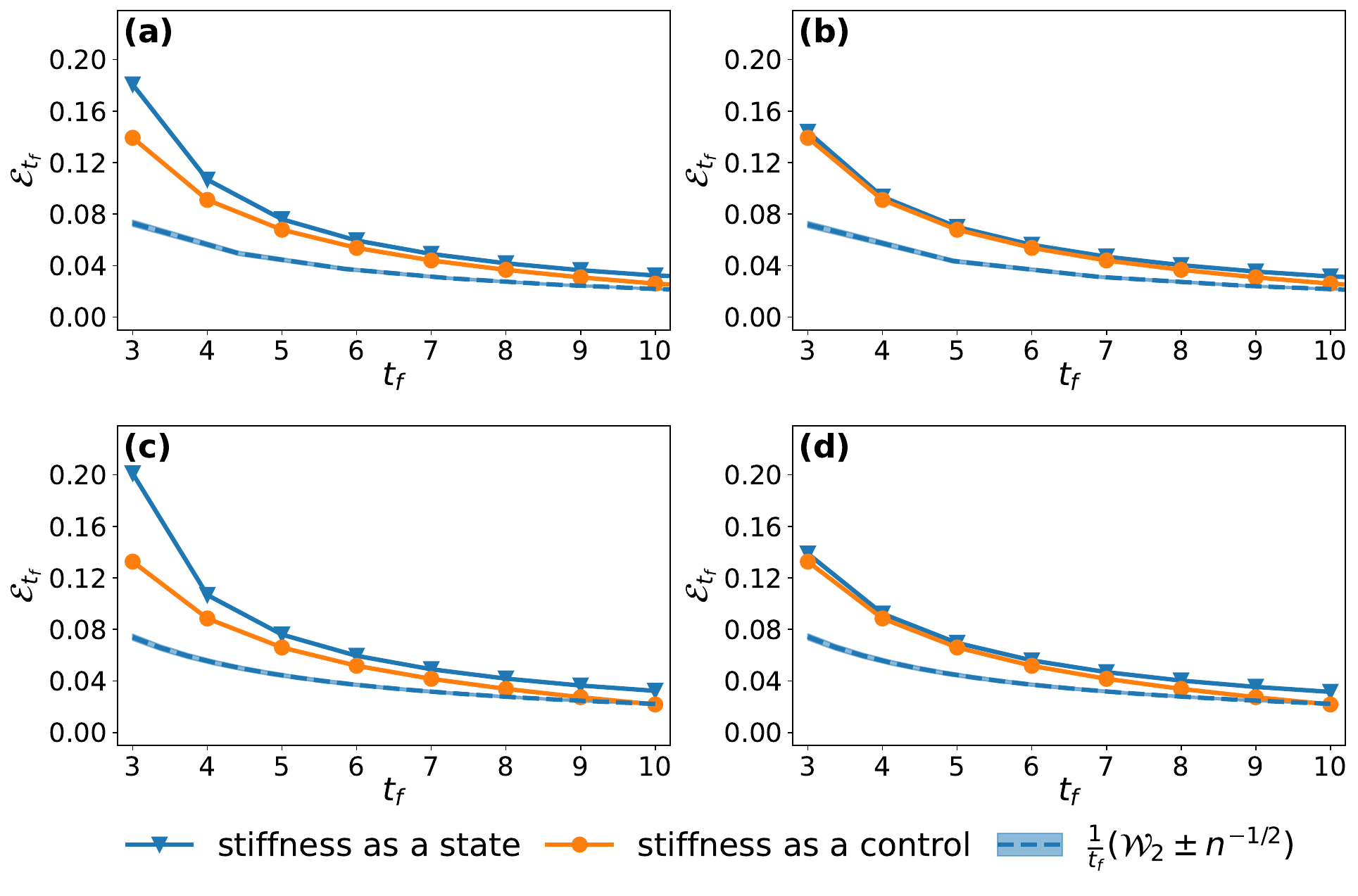}
    \caption{Entropy production as a function of the time horizon $t_f$ where the stiffness is a state \ref{S1} (blue, triangle) and where the stiffness is a control \ref{S2} (orange, circle). When the stiffness is a state, we use a hard penalty $-\Lambda\leq\lambda_t\leq\Lambda$. Panels (a) and (b) constrain the stiffness $\mathscr{k}_{t}$ in the interval $0.2\leq\mathscr{k}_{t}\leq 1.2$. Panels (c) and (d) have the constraint $-0.1\leq\mathscr{k}_{t}\leq 1.2$, allowing for negative values of $\mathscr{k}_{t}$. To model \ref{S2}, only $\mathscr{k}_{t}$ is constrained and there is no constraint on $\lambda_t$ (its time derivative). Results are computed by a direct optimisation of the cost functional. For all plots we use $g=0$ and $\varepsilon=1$; in (a) and (c) we set the size of the hard penalty $\Lambda =1 $; for (b) and (d) we set $\Lambda =10 $. The squared Wasserstein-2 distance $\mathcal{W}_2$ between the initial and final position marginals (blue, dashed) is estimated numerically using $n=20\,000$ independent samples using simulated evolutions of the dynamics controlled by $\mathscr{k}_t$ The blue shaded region indicates the standard error of the estimator. The numerical methods used are as those in Fig.~\ref{fig:harmonic}. 
}
    \label{fig:costs}
\end{figure}

We compare the entropy production as a function of the time horizon in both cases in  Fig.~\ref{fig:costs}. We consider two situations, where we increase the value of $\Lambda$ for the hard penalty, effectively reducing the effect of the penalty and changing the interval where the stiffness $\mathscr{k}_{t}$ is constrained. A larger value of $\Lambda$ is used in panels (b) and (d), where we notice that the difference between the entropy production for cases (\ref{S1}) and (\ref{S2}) becomes smaller in comparison to that shown in panels (a) and (c). In panels (a) and (b), we constrain the value of $\mathscr{k}_{t}$ in the interval $[0.2,1.2]$. This choice of interval is motivated by \cite{BaBePlRaTr2024}. In (c) and (d), we constrain $\mathscr{k}_{t}$ in the interval $[-0.1,1.2]$.

As the final time $\tf$ increases, in both cases the entropy production $\mathcal{E}$~\eqref{td:ep} tends towards the value of the squared Wasserstein distance in configuration space (i.e. position) divided by the time horizon. The squared Wasserstein-2 distance is given by 
\begin{align*}
\mathcal{W}^2_2(\rho_0,\rho_{\tf}) &= \inf_{\pi}\int \mathrm{d} \pi (x,y) \, (x-y)^2     
\end{align*}
using $\rho_0$ and $\rho_{\tf}$ to represent the assigned initial and final position marginals respectively, and $\pi(x,y)$ are joint distributions such that the marginal of $x$ is $\rho_0$ and the marginal of $y$ is $\rho_{\tf}$. The squared Wasserstein-2 distance divided by final time specifies the minimum value of the entropy production in the overdamped (Langevin-Smoluchowski) limit \cite{AuGaMeMoMG2012} and provides a lower bound on the entropy production by the underdamped dynamics \cite{SaBaMG2024}.

\begin{figure}[!h]  
    \centering
    \includegraphics[width=0.9\linewidth]{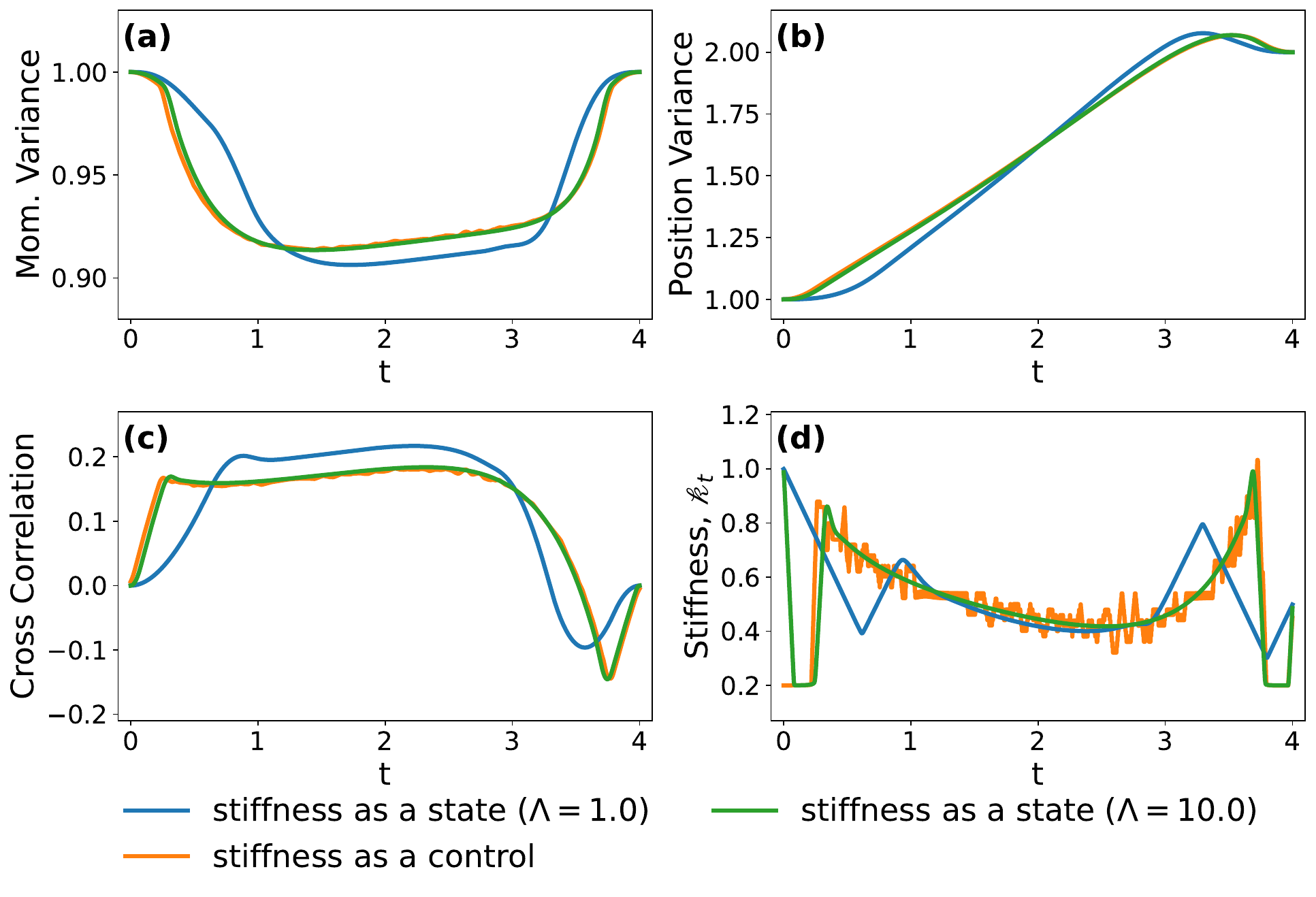}
    \caption{Comparison of engineered swift equilibration at minimum entropy production when the stiffness is the control \ref{S2} (orange) and when the stiffness is a state \ref{S1} (blue, green). The stiffness $\mathscr{k}_{t}$ is constrained in the interval $0.2\leq\mathscr{k}_{t}\leq 1.2$ in both cases. We use a hard penalty~\eqref{penalty:hard} to model the case when the stiffness is a state, with $\Lambda=1$ (blue) and $\Lambda=10$ (green). Results are computed by a direct optimisation (see Fig.~\ref{fig:harmonic}) with parameters $\varepsilon=1$ and $g=0$. We use boundary conditions~\eqref{ep_num:bc} when the stiffness is a state (SI) and remove the boundary conditions on $\mathscr{k}_{t}$ for stiffness as the control (case \ref{S2}). We use $\sigma^2_0 = 1$ and $\sigma^2_{\tf}=2$. Final quantities are smoothed by a convolution with box filter over an interval of approximately $0.067$}
    \label{fig:cumulants_kappa_constrained}
\end{figure}

\begin{figure}[!h]  
    \centering
    \includegraphics[width=0.9\linewidth]{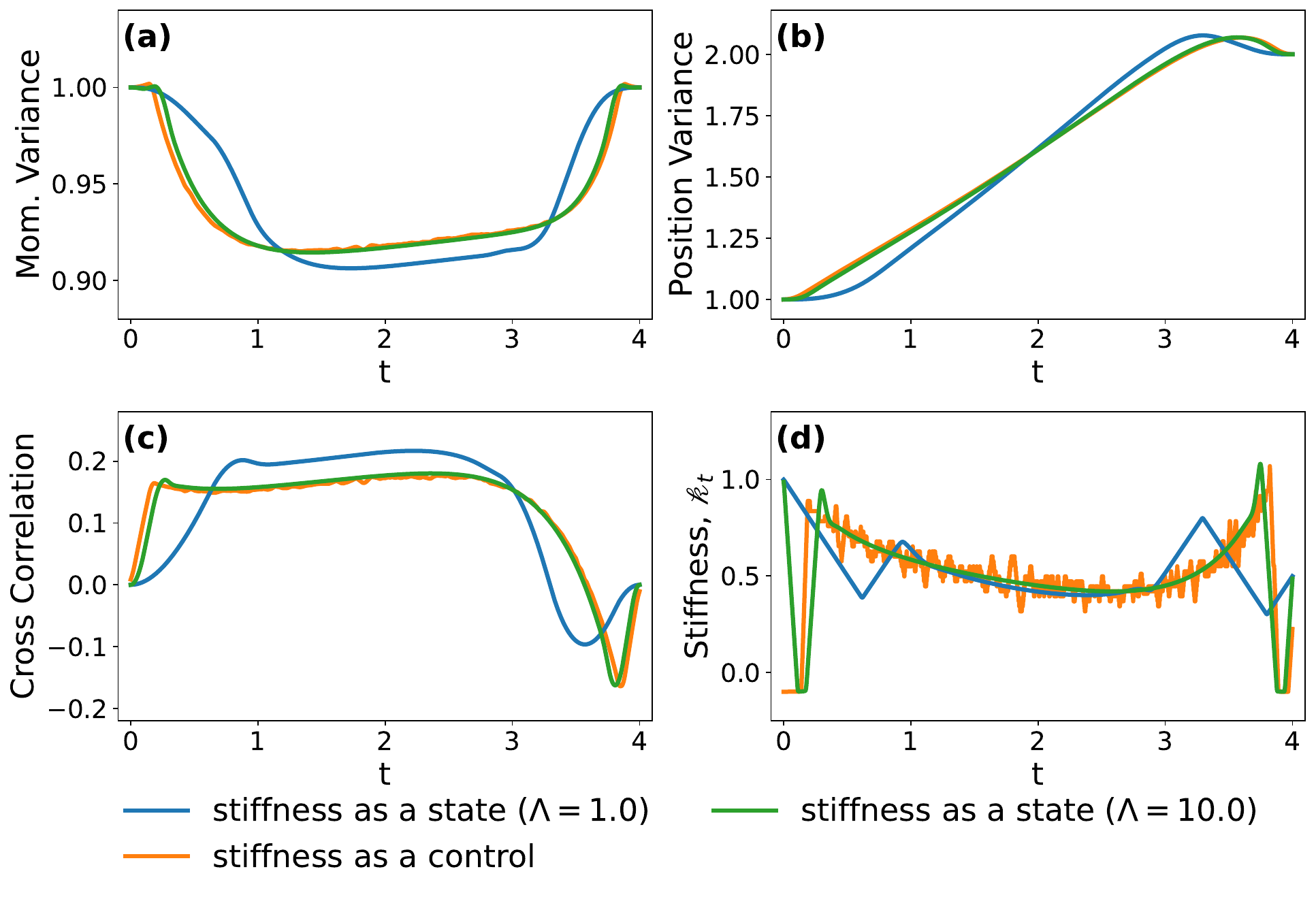}
    \caption{Comparison of engineered swift equilibration at minimum entropy production when the stiffness is the control \ref{S2} (orange) and when the stiffness is a state \ref{S1} (blue, green). The stiffness $\mathscr{k}_{t}$ is constrained in the interval $-0.1\leq\mathscr{k}_{t}\leq 1.2$ in both cases, allowing for negative values. We use a hard penalty~\eqref{penalty:hard} to model the case when the stiffness is a state, with $\Lambda=1$ (blue) and $\Lambda=10$ (green). Results are computed by a direct optimisation (see Fig.~\ref{fig:harmonic}) with parameters and boundary conditions as in Fig.~\ref{fig:cumulants_kappa_constrained_negative}. Final quantities displayed are smoothed by a convolution with box filter over an interval of size approximately $0.067$}
    \label{fig:cumulants_kappa_constrained_negative}
\end{figure}

The cumulants of a transition where the stiffness is a state and a control are shown in Fig.~\ref{fig:cumulants_kappa_constrained} and Fig.~\ref{fig:cumulants_kappa_constrained_negative}. We notice that for stiffness as a control, the results are more noisy for the stiffness. In fact, as the data is computed using a direct method, we obtain a solution that jumps from the limits of the constraining interval of $\mathscr{k}_{t}$, in this case between $0.2$ and $1.2$. Applying a standard mean filter convolution onto the data uncovers the centre manifold, which is in good agreement with case~\ref{S1}, where the stiffness is a state. The direct method is attempting to perform synthesis on the interval, however lacking sufficient accuracy due to discretisation, can only find the turnpike manifold ``on average''. As the value of $\Lambda$ gets larger, i.e. the penalty on $\lambda_t$ becomes weaker, the behaviour of the cumulants becomes more similar to the case where the stiffness is the control. Constraining $\mathscr{k}_{t}$ in a larger interval which also includes negative values is shown in Fig.~\ref{fig:cumulants_kappa_constrained_negative}. From the physical point of view, temporarily negative values of the stiffness of the driving potential are physically realisable \cite{AlLaJu2020} considering that inside the control horizon this quantity is not related to the variance of the Langevin particle position process. Negative values of the stiffness just mean an acceleration pushing the particle away from the origin. 

\subsection{Minimum Work Transitions}
\label{sec:mwt}

Minimum work transitions of the type \ref{C2} minimise the mean work~\eqref{work:ep} in a transition to a target state generically out of equilibrium. Namely, the stationary condition \eqref{noneq_bc} relates the co-state of the stiffness to the position variance. This condition implicitly determines the terminal value of the stiffness minimising the thermodynamic work during the transition. In other words, the terminal value of the mechanical potential, i.e. the internal energy, is not assigned as a boundary condition but derived as an element of the solution of the optimization problem. This is different from the optimal control problem introduced in \cite{ScSe2007} with the interpretation of work minimization and then also studied in \cite{AuMeMG2011}. The optimal control considered in these papers is equivalent to the minimization of the work functional \eqref{mw:cost} when we assign the value of the potential stiffness at the end of the control horizon. Furthermore, if we identify the stiffness as a control as in case~\ref{S2} above, there is no reason why its end-of-horizon value determined by the stationary condition should be equal the value of the quantity interpreted as stiffness in the terminal cost. This is because of the over-determination issue discussed in the introduction. The bottom line is that by formulating work minimization in a proper Bolza form, not only are we avoiding discontinuities of the internal energy, a feat that can always be accomplished by regularization \cite{AuMeMG2012}, but we are also in a position to recognize that the physically relevant boundary conditions are not those specified by the assignment of the final value of the mechanical force.

\begin{figure}[!h]  
    \centering
    \includegraphics[width=0.9\linewidth]{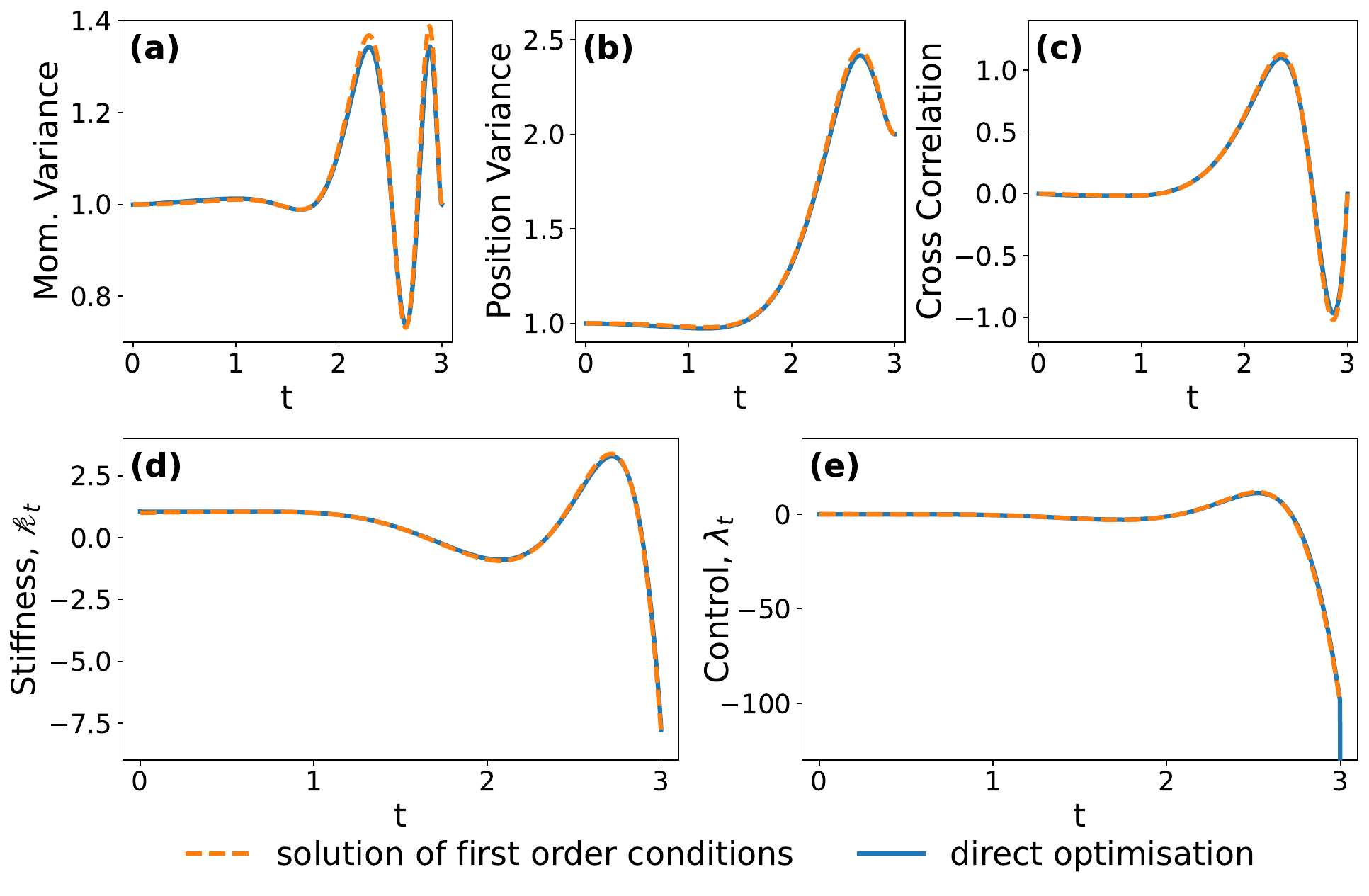}
    \caption{Expansion at minimum work. We use a harmonic penalty and find the solution using a direct optimisation (blue) and as the solution of the first order conditions (orange, dashed). We use $\tf=3,\, \varepsilon=1,\, g=0.01$ and $\Lambda=\sqrt{2}$. We use boundary conditions~\eqref{ep_num:bc} and replace the boundary condition for $x_{\tf}^{(4)}$ with \eqref{noneq_bc}, where $\sigma^2_0 = 1$ and $\sigma^2_{\tf}=2$. Numerical methods are used as those in Fig.~\ref{fig:harmonic}}
    \label{fig:harmonic_noneq}
\end{figure}

\begin{figure}[!h]  
    \centering
    \includegraphics[width=0.9\linewidth]{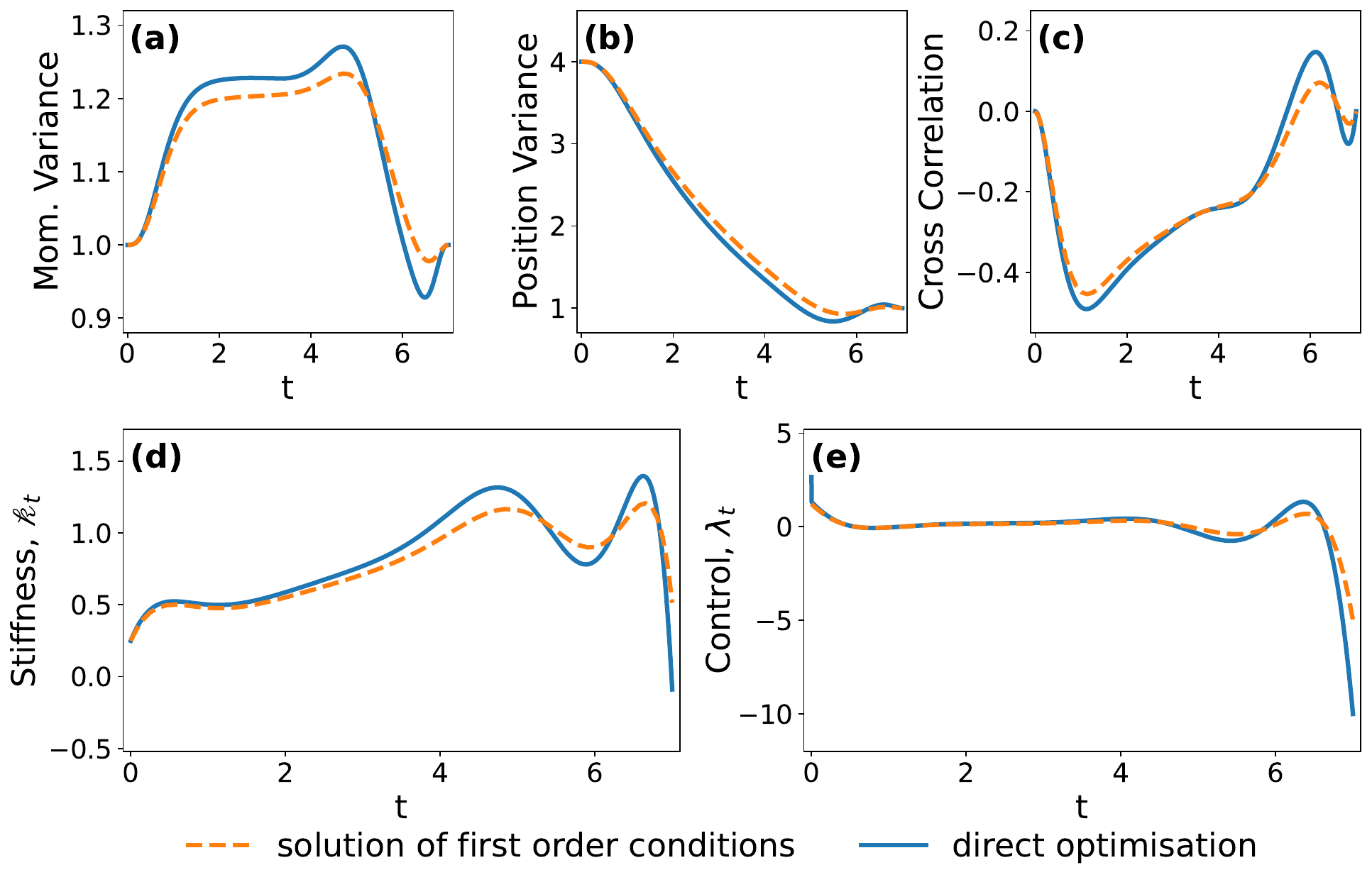}
    \caption{Contraction at minimum work. We use a harmonic penalty and find the solution using a direct optimisation (blue) and as the solution of the first order conditions (orange, dashed). We use $\tf=7,\, \varepsilon=1,\, g=0.1$ and $\Lambda=\sqrt{2}$. We use boundary conditions~\eqref{ep_num:bc} and replace the boundary condition for $x_{\tf}^{(4)}$ with \eqref{noneq_bc}, where $\sigma^2_0 = 4$ and $\sigma^2_{\tf}=1$. Numerical methods are used as those in Fig.~\ref{fig:harmonic}}
    \label{fig:harmonic_contraction_noneq}
\end{figure}
\begin{figure}[!h]  
    \centering
    \includegraphics[width=0.9\linewidth]{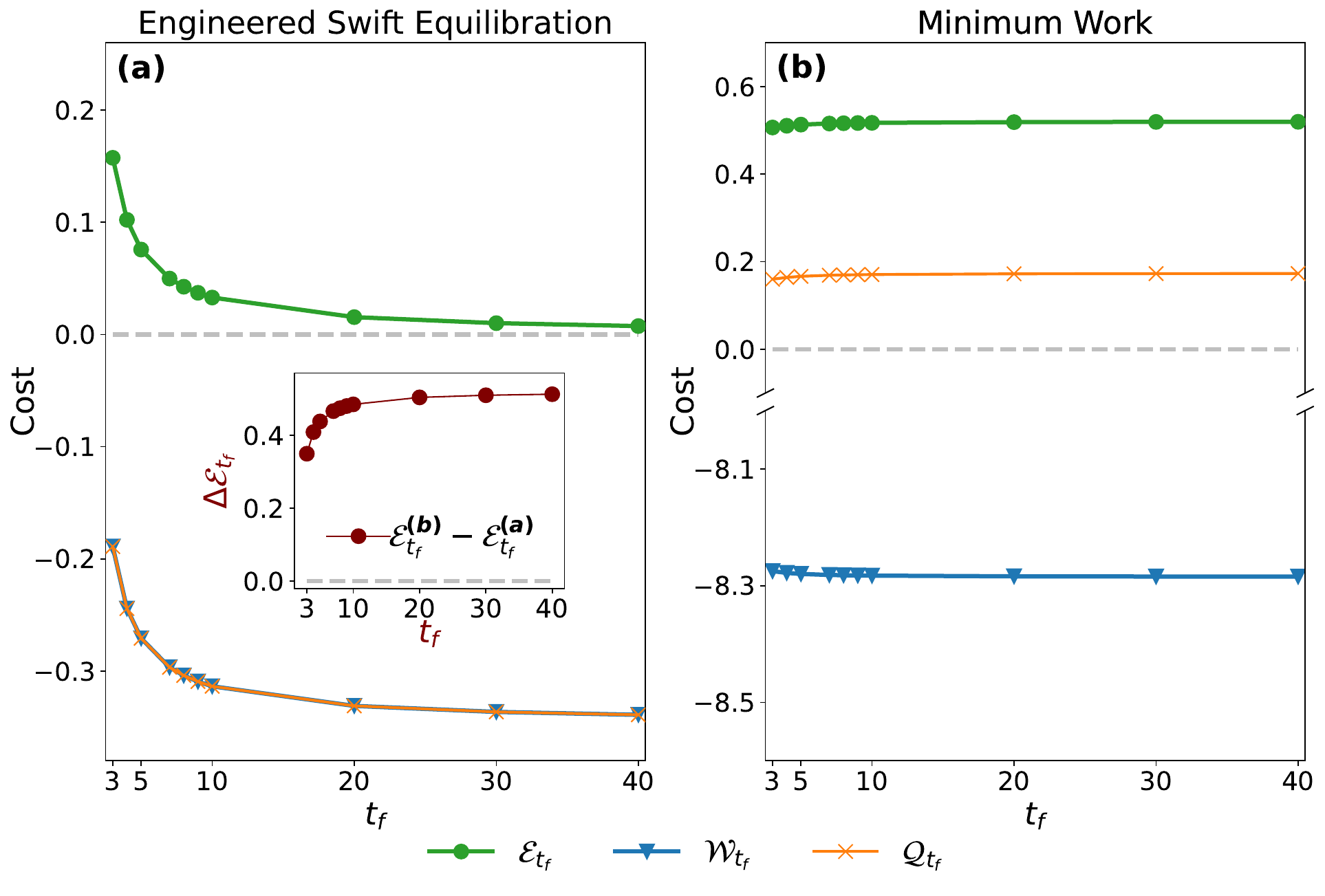}
    \caption{Thermodynamic costs of an expansion for engineered swift equilibration (a) and at minimum work (b). We show the mean work ($\mathcal{W}_{\tf}$~\eqref{td:work}, blue triangle); mean heat release ($\mathcal{Q}_{\tf}$, orange cross); mean entropy production ($\mathcal{E}_{\tf}$~\eqref{td:ep}, green dot) as functions of the time horizon. Inset in panel \textbf{(a)} shows the difference in entropy production between the engineered swift equilibration and minimal work transition, showing that the entropy production is always higher for transitions at minimum work rather than between equilibrium states. Costs are computed using solutions of first order conditions with a harmonic penalty, see Fig.~\ref{fig:harmonic}, with parameters $\varepsilon=1,\, g=0.01$ and $\Lambda=\sqrt{2}$. We use boundary conditions~\eqref{ep_num:bc} in panel \textbf{(a)}; in panel (b) we replace the boundary condition for $x_{\tf}^{(4)}$ with  \eqref{noneq_bc}. We use $\sigma^2_0 = 1$ and $\sigma^2_{\tf}=2$ in both cases.}
    \label{fig:harmonic_costs}
\end{figure}

\begin{figure}[!h]  
    \centering
    \includegraphics[width=0.9\linewidth]{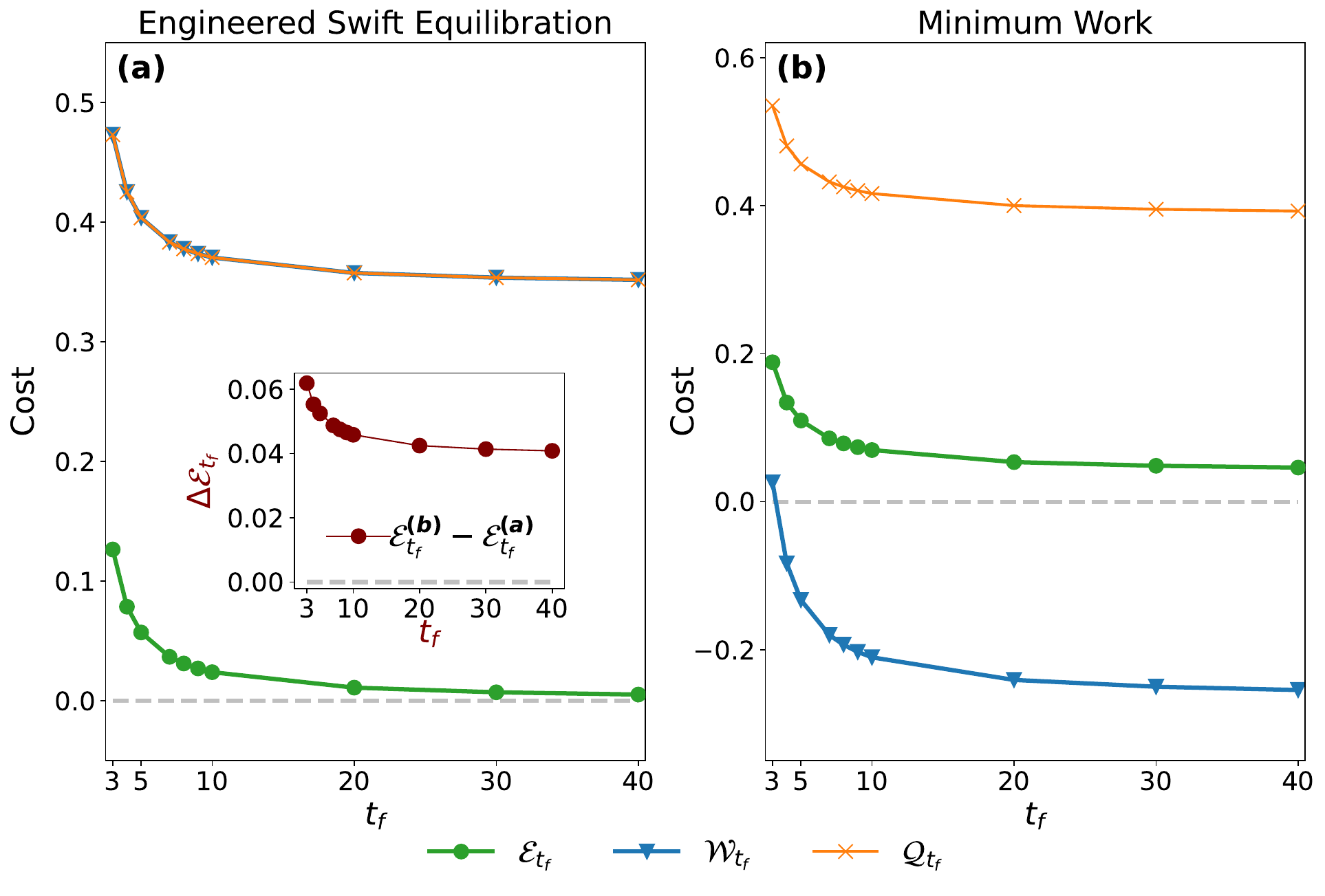}
    \caption{Thermodynamic costs of an expansion for engineered swift equilibration (a) and at minimum work (b). We use parameters and boundary conditions as Fig~\ref{fig:harmonic_costs}, and set $\sigma^2_0 = 1$ and $\sigma^2_{\tf}=0.5$ to model the contraction.}
    \label{fig:harmonic_costs_contraction}
\end{figure}

 We illustrate an expansion at minimum work in Fig.~\ref{fig:harmonic_noneq} and a contraction in Fig.~\ref{fig:harmonic_contraction_noneq}. Minimum work transitions prove to be more numerically unstable than those between equilibria, which leads to small discrepancies between the direct and indirect methods, particularly for the contraction shown in Fig.~\ref{fig:harmonic_contraction_noneq}.

We compare the thermodynamic costs between equilibrium and non-equilibrium transitions: in Fig.~\ref{fig:harmonic_costs} for an expansion, and in Fig.~\ref{fig:harmonic_costs_contraction} for a contraction. In general, entropy production is a measure of how far away a system is from equilibrium. Therefore we expect that engineered swift equilibration should give a smaller entropy production, which is indeed the case in both expansion and contraction (see insets of Figs.~\ref{fig:harmonic_costs} and~\ref{fig:harmonic_costs_contraction}). 

\subsection{Turnpike behaviour}

In all examples, we notice that optimal protocols exhibit a tendency to align with a universal behaviour in the bulk of the control horizon. There, the cumulant dynamics remains close to the values determined by the centre manifold identified in section~\ref{sec:centre}. This universal behaviour becomes more pronounced as $g$, the order parameter of the penalty on the control potential~\eqref{penalty:general}, tends to zero, see Figs~\ref{fig:harmonic} and ~\ref{fig:compactlog_method}. In the case of the hard penalty~\eqref{penalty:hard}, 
this limit corresponds to larger values of the bound $\Lambda$. 
We interpret the existence of a universal regime of the dynamics as the so-called ``turnpike'' phenomenon, which is often observed in optimal control problems. The name turnpike, an American English term for highway, originates in econometric studies of neo-classical models of optimal capital growth such as the Ramsey problem of optimal saving \cite{SamP1965}. In the econometric context, the turnpike property refers to the tendency of time evolution of capital to remain in a neighborhood of a ``bliss'' point specifying the maximum of a strictly concave consumption utility function in the static case. 
An alternative terminology designating the same phenomenon is that of ``von Neumann" path, to honor the groundbreaking work \cite{vNeJ1945} containing one of the very first applications of linear programming to prove the existence of an equilibrium point in neoclassical economic growth models.

More generally \cite{TrZu2015}, the turnpike property stipulates that the qualitative behaviour of optimally controlled dynamics can be subdivided into three regions. Two of them are located at the beginning and the end of the control horizon. There, the optimal dynamics is strongly affected by inherently non-universal constraints imposed by the boundary conditions: the form of the initial state of the system, its final target state, or, alternatively, the presence of a terminal cost and the penalty affecting the choice of controls that is adapted to match these conditions. When the control horizon becomes large, or the intensity of the penalty on controls becomes smaller, in our case  measured by $g$, these regions are compressed to boundary layers, and the behaviour in the bulk converges exponentially to the turnpike. Here, the form of the running cost dominates the dynamics. 

In our set up, the running cost is specified by the momentum variance. The numerical integration of the centre manifold equations (\ref{sf:nonuniv}), (\ref{sf:univ}) in Fig.~\ref{fig:harmonic} shows how the turnpike behaviour corresponds to maintaining the momentum variance close to a constant value near the Maxwell-Boltzmann equilibrium value, equal to unity in our non-dimensional units.

	\section{Conclusions and outlooks}
	
	Optimal control theory problems of the form considered in this paper 
    make a neat distinction between the 
    control horizon (when the system is externally steered between target states) and the time interval when the system evolution is only monitored. 
    This distinction unavoidably implies the existence of non-analytic mathematical indicators that jump from one value to another when the control is switched on or off. The problem is therefore how to properly interpret the role of indicators, when comparing predictions of mathematical optimal control models to experimental results. 
	
	Our interpretation is that thermodynamic quantities at the end of an optimal control horizon are only unambiguously defined when they are pure functions of the system's state variables. In agreement with standard formulations of the theory \cite{LibD2012}, excluding an explicit dependence on control protocols avoids identifying, by construction, terminal costs with markers of intrinsic discontinuity in mathematical models. For instance, the work done on the system is well defined within an optimal control model if the internal energy is a (pure function of the) system's state variable at the beginning and end of the control horizon. This motivates 
    a dynamic model for the evolution of the mechanical potential acting on the system. 
    In doing so, we also address the issue that in actual laboratory implementations, control protocols cannot be switched on and off instantaneously; ignoring this may lead to experimental setups that incur in unwanted consequences.  For instance, \cite{HiCa2018} warns that neglecting the finite response time of reactive elements of tuned circuits for resonant driving of electric and magnetic fields causes to signal distortions.

	In addressing these difficulties, we 
    need to consider higher order Markov decision processes, where the thermodynamic quantities do not specify a running cost that is convex in the controls. A solution with straightforward physical justification \cite{BaBePlRaTr2024} is to restrict admissible controls to take values on a compact set. Mathematically, the resulting formulation requires the construction of optimal controls via a synthesis procedure \cite{BoPi2005}. Actual computation of synthesised trajectories may, however, become challenging. Based on the theory of self-concordant functions in direct optimal control methods \cite{KarN1984,NeNe1994,NeTo2008}, we demonstrate that, for practical purposes, modifying thermodynamic cost functionals to include a convex penalty on controls
    , produces qualitatively and, often, quantitatively equivalent results. 
    
    This can be explained in two ways. From the numerical point of view, the most efficient available algorithms for direct optimisation actually model confinement in a compact set by means of self-concordant smooth barriers \cite{WacA2009}. From the analytic point of view, we demonstrate by asymptotic analysis that, independently of the penalty imposed on admissible protocols, in the bulk of the control horizon optimal evolution tends to remain close to a universal centre manifold, described by fewer  
    degrees of freedom. The system then  
    deviates from the centre manifold only at the beginning and end of the horizon, to match boundary conditions.

    This behaviour is the turnpike phenomenon often observed in optimal control \cite{TrZu2015}. The phenomenon explains the observed thermodynamic efficiency of heuristic protocols in engineered swift equilibration \cite{ChCiGuOdTr2018,DaCiBe2023}. It also clarifies, in our view conclusively, that the predictive power of models identifying the control with the microscopic potential (see e.g. \cite{SaBaMG2024} and references within) indeed refers to the turnpike or bulk time interval. The idea of optimal protocols with jumps at the boundaries should be discarded as a potentially misleading metaphor for unresolved time scales in a model.
	
	In the present paper we restricted our study to Gaussian models with linear dynamics. Similar ideas can be also applied to feedback control for transitions between states described by  more general distributions in phase space, such as those entering Landauer's model of bit erasure \cite{AuGaMeMoMG2012}. We plan this extension to be the subject of forthcoming work.

	\section{Acknowledgments}
	
	The authors gratefully acknowledge Marco Baldovin, Dario Lucente and Jukka Pekola for many discussions and insights. The authors wish to acknowledge CSC – IT Center for Science, Finland, for computational resources. JS acknowledges financial support from the Doctoral Programme in Mathematics and Statistics at the University of Helsinki and the centre of Excellence FiRST of the Research Council of Finland (funding decision number: 346305).
	
    \appendix
    \setcounter{section}{0}    \section{Absence of abnormal extremals}\label{sec:abnormal}
	
	Abnormal extremals are non-trivial stationary conditions occurring if we set $$\mathscr{y}_{t}^{\smi{0}}=0$$ in (\ref{MP:pH}). If we now undertake the analysis of section~\ref{sec:synthesis} under this additional condition, we verify that (\ref{noact:c2}) is not affected, and (\ref{noact:c3}) becomes
	\begin{align}
		\mathscr{y}_{t}^{\smi{1}}=\frac{\mathscr{y}_{t}^{\smi{3}}\,\mathscr{x}_{t}^{\smi{3}}}{\mathscr{x}_{t}^{\smi{1}}}
		\nonumber
	\end{align}   
	whence instead of (\ref{noact:a4}), we only arrive at 
	\begin{align}
		0=2\,\varepsilon\,\mathscr{x}_{t}^{\smi{1}}\,\mathscr{y}_{t}^{\smi{3}}
		\nonumber
	\end{align}
	which yields an empty condition on the controls. We thus conclude as in \cite{MGSc2017,BaBePlRaTr2024} that the thermodynamic cost minimisation cannot be obtained by means of abnormal extremals.
	
\section*{References}	
	\bibliography{nanooscillator_control}{} 
    \bibliographystyle{iopart-num}
\end{document}